\documentclass[review]{elsarticle}

\usepackage{hyperref}
\PassOptionsToPackage{table,xcdraw}{xcolor}

\usepackage[table]{xcolor}  
\usepackage{booktabs}
\usepackage{subfigure}
\usepackage{pgfplots}
\pgfplotsset{width=5.5cm}

\usepackage{balance}  
\usepackage{graphics} 

\usepackage{color}
\usepackage{booktabs}
\usepackage{ccicons}
\usepackage{longtable}

\usepackage{url}
\usepackage{fancybox}
\usepackage{multirow}
\usepackage{flushend}
\usepackage{booktabs}
\usepackage{tabularx}
\usepackage{comment}
\usepackage{array}
\usepackage[flushleft]{threeparttable}
\usepackage{mdframed}
\graphicspath{{}{images/}{dia/}}
\DeclareGraphicsExtensions{.pdf,.png,.svg}

\usepackage{listings}
\usepackage{courier}
\usepackage{hyperref}

\usepackage{xpatch}

\xpatchcmd{\refstepcounter}{%
  \stepcounter{#1}%
}{%
  \stepcounter{#1}%
  \xdef\scr{\number\value{#1}}%
}{\typeout{success}}{\typeout{failure}}

\newcounter{o}
\usepackage{tikz}
\definecolor{1c1}{RGB}{188,162,6}
\definecolor{1c2}{RGB}{137,129,80}
\definecolor{1c3}{RGB}{239,167,31}
\definecolor{1c4}{RGB}{88,194,241}
\definecolor{1c5}{RGB}{6,180,188}

\tikzset{mynode/.style={draw=white,solid,circle,fill=green,inner sep=1pt, thick,
text=black}}
\tikzset{arrow line/.style={dashed, line width= 2.5pt, color=#1}}
\usepackage{varwidth}
\def\bf{\textbf}

\def\fig {Figure~}

\def\tbl {Table~}

\def\sec {Section~}

\def\it{\textit}

\usepackage{paralist}

\newcommand{\nd}{\vspace{1mm}\noindent}
\usepackage{tikz}

\lstdefinestyle{inlinecode}{basicstyle={\ttfamily\scriptsize\bfseries}}

\newcommand{\urls}[1]{{\scriptsize\url{#1}}}
\usepackage{tcolorbox}
\newcommand{\emt}[1]{\emph{``#1''}}

\usepackage{paralist}
\usepackage[outercaption]{sidecap}
\usepackage [autostyle, english = american]{csquotes}
\MakeOuterQuote{"}
\newcounter{scn}
\usepackage[shortlabels]{enumitem}
\usepackage{tikz}
\usepackage{local-pgf-pie}
\usepackage{paralist}
\usepackage{soul}

\usepackage{tabularx}

\usetikzlibrary{positioning,shadows}
\usepackage{balance}
\newcommand{\dq}[1]{\href{https://stackoverflow.com/questions/#1/}{$Q_{#1}$}}

\newif\ifpienumberinlegend
\pgfkeys{/number in legend/.code=
    \expandafter\let\expandafter\ifpienumberinlegend
    \csname if#1\endcsname
    \ifpienumberinlegend

    \def\beforenumber##1\afternumber{}%
    \fi,
    /number in legend/.default=true
}
\definecolor{1c1}{RGB}{188,162,6}
\definecolor{1c2}{RGB}{137,129,80}
\definecolor{1c3}{RGB}{239,167,31}
\definecolor{1c4}{RGB}{88,194,241}
\definecolor{1c5}{RGB}{6,180,188}

\tikzset{mynode/.style={draw=white,solid,circle,fill=green,inner sep=1pt, thick,
text=black}}
\tikzset{arrow line/.style={dashed, line width= 2.5pt, color=#1}}

\usepackage{bchart}

\def\test#1{%
    \ifnum #1 > 0
      #1
    \fi
}

\newcommand{\difficultyfivebars}[5]{
{{\color{green}\rule{#1pt}{4pt}} \test{#1}}
{{\color{magenta}\rule{#2pt}{4pt}} \test{#2}}
{{\color{orange}\rule{#3pt}{4pt}} \test{#3}}
{{\color{red}\rule{#4pt}{4pt}} \test{#4}}
{{\color{black}\rule{#5pt}{4pt}} \test{#5}}
}

\journal{Journal of Information and Software Technology}

\begin{document}

\begin{frontmatter}

\title{A Mixed Method Study of DevOps Challenges}

\author{Minaoar Hossain Tanzil\fnref{uofc}\corref{correspondingauthor}}
\ead{minaoar@gmail.com}
\author{Masud Sarker\fnref{buet}}
\author{Gias Uddin\fnref{uofc}}
\author{Anindya Iqbal\fnref{buet}}

\fntext[uofc]{University of Calgary}
\fntext[buet]{Bangladesh University of Engineering and Technology}


\cortext[correspondingauthor]{Corresponding author}

\begin{abstract}
\textbf{Context:} DevOps practices combine software development and IT (Information Technology) operations. The continuous needs for rapid but quality software development requires the adoption of high-quality DevOps tools. There is a growing number of DevOps related posts in popular online developer forum Stack Overflow (SO). While previous research analyzed SO posts related to build/release engineering, we are aware of no research that specifically focused on DevOps related discussions. \textbf{Objective:} This paper aims to learn the challenges developers face while using the currently available DevOps tools and techniques along with the organizational challenges in DevOps practices.  \textbf{Method:} We conduct an empirical study by applying topic modeling on 174K SO posts that contain DevOps discussions. We then validate and extend the empirical study findings with a survey of 21 professional DevOps practitioners. \textbf{Results:} We find that: \begin{inparaenum}[(1)]
\item There are 23 DevOps topics grouped into four categories: Cloud \& CI/CD Tools, Infrastructure as Code, Container \& Orchestration, and Quality Assurance. 
\item The topic category `Cloud \& CI/CD Tools' contains the highest number of topics (10) which cover 48.6\% of all questions in our dataset, followed by the category Infrastructure as Code (28.9\%). 
\item The file management is the most popular topic followed by Jenkins Pipeline, while infrastructural Exception Handling and Jenkins Distributed Architecture are the most difficult topics (with least accepted answers).
\item In the survey, developers mention that it requires hands-on experience before current DevOps tools can be considered easy. They raised the needs for better documentation and learning resources to learn the rapidly changing DevOps tools and techniques. Practitioners also emphasized on the formal training approach by the organizations for DevOps skill development. \textbf{Conclusion:} Architects and managers can use the findings of this research to adopt appropriate DevOps technologies, and organizations can design tool or process specific DevOps training programs.   
\end{inparaenum}
\end{abstract}


\end{frontmatter}

\section{Introduction}\label{sec:intro}
The term `DevOps' was coined in early 2000 to combine software development and IT (Information Technology) practices with the aim to improve software product development and delivery~\cite{Bass-DevOpsSoftwareArch-2015,Loukides-WhatisDevops-2012}. According to UpGuard~\cite{upguard}, 63\% organizations in 2021 experienced  improvement in the quality of their software deployment and frequency of new software releases and 55\% of the organizations noticed improved collaboration among teams  by adopting DevOps tools. Several research studies~\cite{Lwakatare-DevOpsInPractice-IST2019,Roche-DevOpsPracticesQA-ACM2013} show that DevOps adoption can improve software quality assurance and help software teams achieve high degree ownership, but such adoption requires steep learning curve for software developers, operation engineers, and the organization as a whole. As found in the 2015 State of the DevOps report \cite{website:devops-state-2015}, DevOps initiatives launched solely by C-level executives or from the grassroots are less likely to succeed, it is important to understand the challenges engineers face while using DevOps tools and practices for a successful DevOps adaptation.

We observe that discussions about various DevOps problems are prevalent in the popular online developer forum Stack Overflow (SO)~\cite{website:stackoverflow}. Several research has been conducted to
analyze SO posts (e.g., big
data~\cite{bagherzadeh2019going}, concurrency~\cite{ahmed2018concurrency}, blockchain~\cite{wan2019discussed}, microservices~\cite{bandeira2019we}). Recently, SO discussions on modern release engineering are analyzed using topic modeling~\cite{openja2020analysis}. While release engineering practices can belong to DevOps practices, the concept of DevOps is broad as it contains tools and practices around several areas like continuous development, deployment, monitoring, maintenance, etc.
As such, we are aware of no research that analyzed DevOps discussions on SO, 
although such insight can complement existing DevOps literature which so far has mainly used controlled case studies/surveys in the industry~\cite{Lwakatare-DevOpsInPractice-IST2019, Roche-DevOpsPracticesQA-ACM2013, Leite-DevopsSurvey-ACM2020}. 

In this paper, we report a mixed method study to understand the challenges DevOps practitioners face. 
\bf{\ul{First}}, we conduct an empirical study by applying topic modeling on 174K SO posts related to DevOps discussion (\sec\ref{sec:study-results}). We find 23 topics grouped into four categories: Cloud \& CI/CD Tools, Infrastructure as Code, Container \& Orchestration, and Quality Assurance. The category `Cloud \& CI/CD Tools' contains the highest number of topics (10) and the most number of questions ($>$48\%), followed by the category `Infrastructure as Code' with 7 topics and 28.9\% of all questions. File management and Jenkins pipeline are the most popular (with most views) topics, while Test Automation and Git CI/CD support are the most difficult topics with highest percentage of questions without an accepted answer.
\bf{\ul{Second}}, we validate and extend the findings of the empirical study using a survey of 21 professional DevOps practitioners (\sec\ref{sec:survey}). We find that the survey participants agreed with our observed topics of DevOps challenges in SO. We also find that organizational training is insufficient to improve DevOps skills. Moreover, experts shared difficulties in version compatibility, and strongly predicted that in future, cloud infrastructure automation will be future trend in DevOps. Our research contribution is summarized in Table \ref{tab:contribution_summary}.

\begin{table}[]
    \centering
        \caption{Summary of research contributions and advancements made by our study}
    \resizebox{\columnwidth}{!}
    {
    \begin{tabular}{p{1.5cm}|p{5.5cm}|p{11cm}}
        \textbf{Method} & \textbf{Research Contribution} & \textbf{Research Advancement} \\ \toprule
         Empirical Study 
         
         & Out of 174K SO posts, we observed 23 DevOps topics under 4 major categories of Cloud \& CI/CD Tools, Infrastructure as Code, Container \& Orchestration, and Quality Assurance. We also analyzed most viewed (File management and Jenkins pipeline) and most unresolved challenging topics (Test Automation and Git CI/CD support) in SO. 
         
         & As per our knowledge, no other study analyzed DevOps topics in SO. Previous studies in closely related domain of continuous software engineering and release engineering did not find major DevOps topics that we found, e.g., Kubernetes, Cloud-Infra Automation, Non-functional Test Automation etc. Our study found that Kubernetes, which is a very important part of modern infrastructure orchestration, comprises of 13.4\% of all DevOps posts; Cloud-Infra Automation has 6.2\% posts; and Non-function Test Automation related posts take the highest amount of time (251 hours) to get a response among all topics. Moreover, no other previous researches found staggering 21.4\% of DevOps posts related with Jenkins only. These unique findings provide significant insights to SE researchers and infrastructure and test automation vendors. \\ \hline
         
         Survey 
         
         & We conducted a practitioners survey to evaluate our empirical findings and to collect people and process related DevOps challenges. 21 DevOps practitioners responded to quantitative and open-ended questions on DevOps tools, practices, challenges, and trends.
         
         & We reported that DevOps related skill shortage is major bottleneck faced by practitioners in the industry. Still only 5\% of organizations provide formal training for DevOps. As we found prevalence of Cloud-Infra Automation topic in our empirical study, practitioners also re-enforced strong prediction in this future technology trend. These kind of qualitative insights related with people, organization, and technology trend are critical for management and improvement of DevOps practices which would go unnoticed otherwise.
         \\ \hline

         Tag set analysis
         
         & At the initial stage of our empirical study, we had to select a final tag-set from 30 different tag-sets. We provided detailed replicable steps how such topic related final tag-set can be generated through manual analysis.
         
         & Though it is common in empirical studies to generate a list of relevant tag-sets using some significance and relevance values, we could not find any other study to explain how they choose the final tag-set among dozens of such tag-sets. We have provided step-by-step detail analysis process with examples on how we discarded the non-relevant tags and finally selected the relevant tag-set. Future empirical studies can be significantly benefited to replicate this method. \\
         
         \bottomrule
    \end{tabular}
    }
    \label{tab:contribution_summary}
\end{table}

Our study findings can be useful to several stakeholders in Software Engineering (SE). The DevOps developers and practitioners can use the observed topics to prioritize/improve their learning of DevOps tools. DevOps vendors can guide their tool development and SE researchers can develop innovative techniques and documentation to address the challenges.

\begin{figure}
    \centering
    \includegraphics[scale=.75]{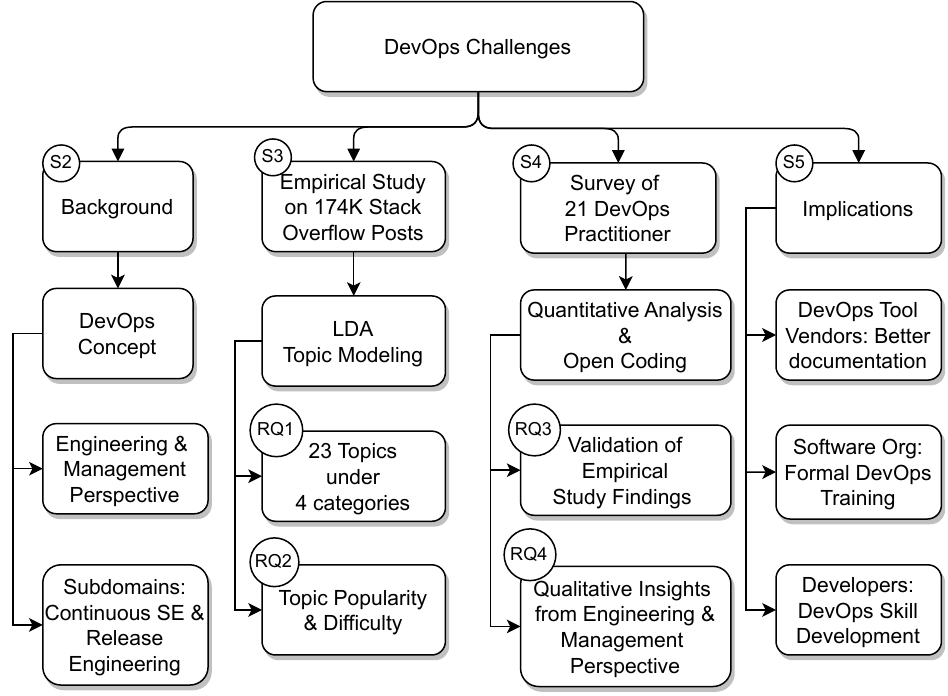}
    \caption{Outline of the paper, describing four major sections and four research questions. S-n refers to section n, and RQ-n refers to search question n.}
    \label{fig:outline}
\end{figure}

The paper is outlined as follows: we explore the DevOps concepts and related subdomains in section 2, we present the methodology of empirical study performed on Stack Overflow and also the topics and difficulties found in section 3 which is followed by a practitioners' survey design and findings that can validate and extend the empirical results. Finally, implications for the industry and academia in section 5 is followed by threats to validity and related works in section 6 and 7 respectively. The total flow of the paper is outlined in Figure \ref{fig:outline}.

\nd\bf{Replication Package:} \url{https://github.com/DevOpsTopic/Challenges} \cite{replciation-package}.

\section{Background}\label{sec:background}
\begin{figure}
    \centering
    \includegraphics[scale=0.5]{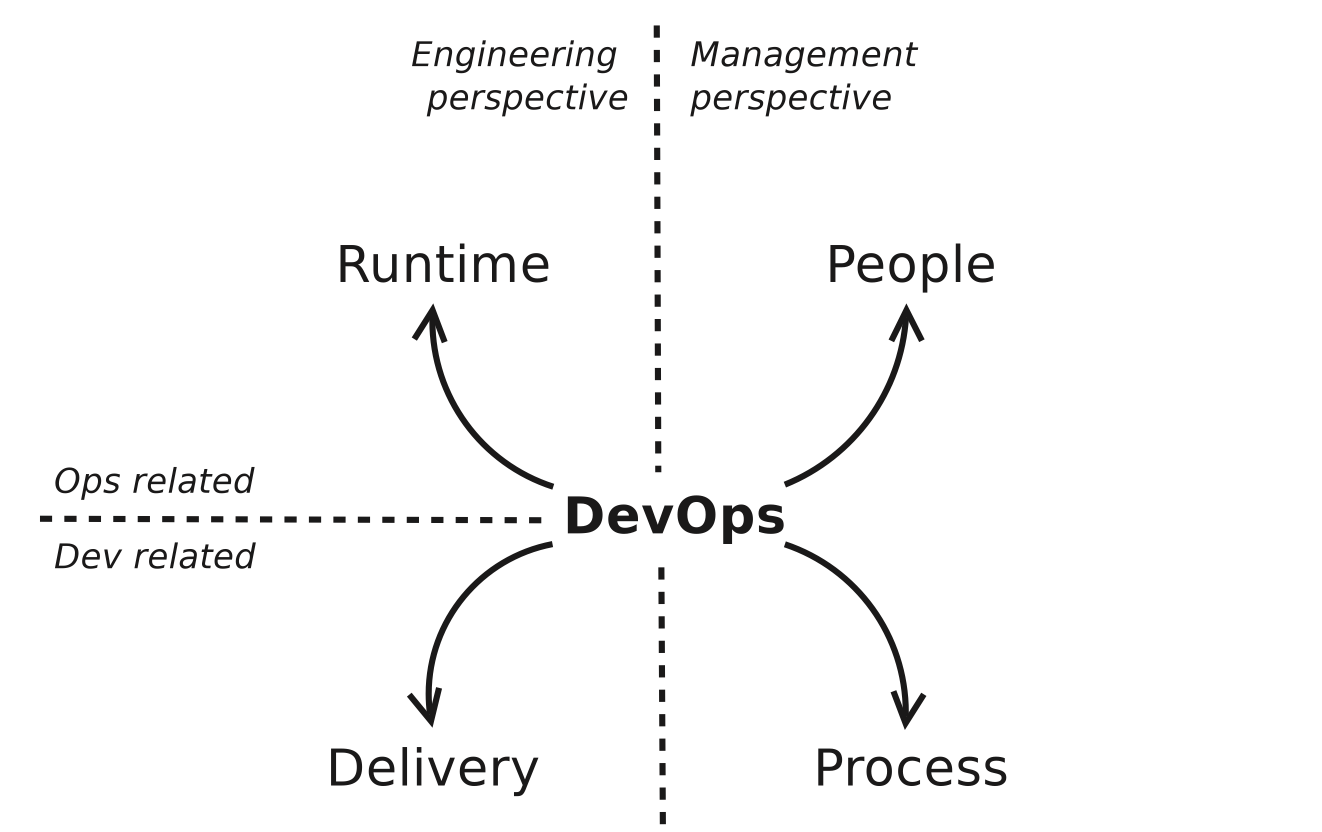}
    \caption{DevOps Conceptual Map provided by Leite er al. \cite{Leite-DevopsSurvey-ACM2020}}
    \label{fig:concept_map}
\end{figure}

\bf{DevOps Concepts.} Aiello et al. defined DevOps in \cite{Aiello-AgileDevOps-2016} as \textit{DevOps is a set of principles and practices intended to help development and operations collaborate and communicate more effectively.} A slightly extended definition is provided by Leite et al.~\cite{Leite-DevopsSurvey-ACM2020}, DevOps \emt{is a collaborative and multidisciplinary organizational effort to automate continuous delivery of new software updates while guaranteeing their correctness and reliability}. In this literature survey, Leite et al. \cite{Leite-DevopsSurvey-ACM2020} identified four major categories of DevOps concepts, namely, Process, People, Delivery, and Runtime using a systematic analysis on 50 'core' DevOps publications till 2019. While process and people are relevant from the management perspective, delivery (development related) and runtime (operation related) are associated from the engineering perspective. The categories are depicted in Figure \ref{fig:concept_map}.

The \textbf{process} category covers business-related planning and strategies. \textbf{People} encompasses concepts regarding skills, incentives and the culture of collaboration. \textbf{Delivery} consists of the concepts necessary for continuous build, delivery and releases, and, finally, \textbf{Runtime} elaborates concepts related to the guarantee of operational stability and reliability of services in a continuous software engineering environment.

\bf{Continuous Software Engineering Processes in DevOps.} 

Given the broad conceptual coverage of DevOps, it is closely related with the processes of Continuous Software Engineering (CSE). Fitzgerald et al. extended the concept of CSE and proposed a holistic approach of 'Continuous *' in \cite{fitzgerald2014continuous}. They divided the entire software life-cycle into three main sub-phases: Business Strategy and Planning, Development, and Operations. Within these sub-phases, they positioned the various categories of continuous software engineering activities (all-continuous planning, integration, deployment, delivery, verification, testing, compliance, monitoring etc.). Finally, they positioned continuous improvement and innovation as the foundation upon which other 'Continuous *' activities could be grounded. Comparing this proposition of CSE by Fitzgerald et al. \cite{fitzgerald2014continuous} and DevOps conceptual framework by Leite et al. \cite{Leite-DevopsSurvey-ACM2020}, CSE covers three concepts of DevOps, namely Process, Delivery, and Runtime and does not consider the People category of DevOps.

\bf{Release Engineering in DevOps.} As noted by Dyck et al. \cite{releng-devops-7169442}, \textit{release engineering and DevOps terms are often confused, misinterpreted, or used as synonyms.} In their paper, they proposed definitions of Release Engineering (RE) and DevOps to keep them clearly distinct. Adams and McIntosh \cite{releng-adams-2016} defined RE as \textit{The release engineering process is the process that brings high quality code changes from a developer’s workspace to the end user, encompassing code change integration, continuous integration, build system specifications, infrastructure-as-code, deployment and release.} With this definition, RE is equivalent to the Delivery category of DevOps as explained by Leite et al. \cite{Leite-DevopsSurvey-ACM2020}.

\bf{Relation among DevOps, Continuous Software Engineering and Release Engineering.} To summarize these closely related but often mixed up concepts, we refer to the Table \ref{tab:devops_cse_re} with the definition and activities of each concept outlined by the respective researches mentioned above. 

\begin{table}[]
    \centering
    \caption{Relation and distinction among DevOps, Continuous Software Engineering (CSE), and Release Engineering (RE).}
    \resizebox{\columnwidth}{!}
    {
    \begin{tabular}{l|l|l}
        \textbf{DevOps} & \textbf{CSE} & \textbf{Release Engineering} \\ 
        (Leite et al. \cite{Leite-DevopsSurvey-ACM2020}) & (Fitzgerald et al. \cite{fitzgerald2014continuous}) & (Dyck et al. \cite{releng-devops-7169442}) \\ \toprule \hline
         \multirow{3}{*}{Process} & Continuous Planning & \\ \cmidrule{2-2}
         {} & Continuous Improvement & \\
         {} & Continuous Innovation & \\ \hline \hline
         \multirow{8}{*}{Delivery} & Continuous Integration & Continuous integration: Building \& Testing \\
         {} & Continuous Testing & Integration: Branching and Merging \\ 
         {} & Continuous Verification & Build System \\ \cmidrule{2-3}
         {} & Continuous Deployment & Deployment \\ \cmidrule{2-3}
         {} & \multirow{2}{*}{Continuous Delivery} & Infrastructure-as-Code \\ 
         {} & {} & Release \\ \cmidrule{2-3}
         {} & Continuous Compliance & \\ \cmidrule{2-2}
         {} & Continuous Security & \\ \cmidrule{1-3} \hline
         \multirow{4}{*}{Runtime} & Continuous Use & \\
         {} & Continuous Trust & \\
         {} & Continuous Run-time & \\
         {} & Continuous Monitoring \\ \cmidrule{1-3} \hline
         People & & \\ \bottomrule
    \end{tabular}
    }
    \label{tab:devops_cse_re}
\end{table}

\begin{figure}
    \centering
    \includegraphics[width=1\textwidth]{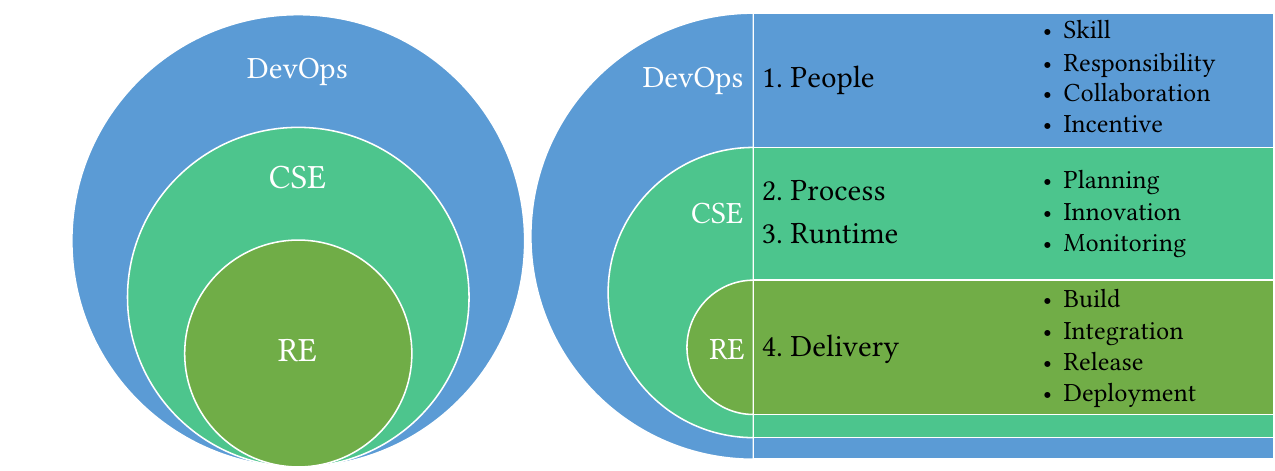}
    \caption{Superset relation of DevOps, CSE, and RE}
    \label{fig:devops-cse-re}
\end{figure}

From the detailed activities, definition, and framework of DevOps, CSE, and RE, it is evident that DevOps is a superset of CSE which is a superset of RE. This final relation is depicted in the Figure \ref{fig:devops-cse-re}.

\section{Empirical Study of DevOps Discussion}\label{sec:study-results}
In this section, we apply topic modeling on 174K SO posts related to DevOps discussion and answer two research questions:
\begin{enumerate}[label=\bf{RQ\arabic{*}.}]
  \item What types of topics are discussed about DevOps in SO?
  \item How do the popularity and difficulty of DevOps topics vary?
\end{enumerate} As DevOps is a fast evolving complex process of tools and culture, it is important to recognize the vocabulary and topics in DevOps. In RQ1, we analyze DevOps topics in SO discussions to inform of the challenges around DevOps tools and techniques. All the DevOps topics we observed in RQ1 may not be equally popular or equally difficult. In RQ2, we study of popularity and difficulty of the topics to offer insight into prioritizing research and organizational effort. For example, system architects may decide to choose matured tools with less difficulty or cutting edge tools with active community support \cite{van2010importance}, organizations may decide which topics need extra investment in terms of employee training. Software engineering researchers and DevOps vendors could also devise ways to make improved tools and techniques using the knowledge. 

\subsection{Study Data}

We collect SO posts in three steps: (1) Download SO dataset, (2) Develop DevOps tag set, and (3) Extract posts using tag set. 

\subsubsection{\textbf{Download SO Dataset}}
We downloaded the SO dataset which is publicly available in SOTorrent ~\cite{website:stackoverflow-dataset}. The dataset includes SO question and answer posts and their metadata. The metadata includes post identifier, its type (question or answer), title, body, tags, creation date, view count, score, favorite count, and the identifier of the accepted answer (if the post is a question). The collected dataset includes 49,139,907 questions and answers posted over the last 13 years from August 2008 to June 2020.

\subsubsection{\textbf{Develop DevOps Tag Set}}

In this step we develop a set of DevOps tags from a initial tag \(T_{init}\)= {DevOps}. Then from our dataset (Q) we extract the question posts (P) whose tags match a tag in \(T_{init}\). From the initial posts (P), we extract the candidate tag set (t). The candidate tags are evaluated by three criteria A, B, and C. In particular, 'A' denotes the number of posts from P for each of the candidate tags t (number of posts of questions whose tag contains t and "DevOps"). 'B' denotes the number of all posts in Q for each of the candidate tags t (number of posts in the whole dataset for each of the candidate tags t). Additionally, 'C' denotes the number of posts in P. Tags are then refined further by their relevance. Two heuristics (µ) and (v) are used to measure from previous work ~\cite{yang2016security,Ahmed-ConcurrencyTopic-ESEM2018,Bagherzadeh-BigdataTopic-FSE2019} and relevance of a candidate tag t in the DevOps tag set (t).

\[
Significance (\mu) = \frac{(A)\, Number\, of\, posts\, with\, tag\, t\, in\, P}{(B)\, Number\, of\, posts\, with\, tag\, t\, in\, Q }
\]
\[
Relevance (v) = \frac{(A)\, Number\, of\, posts\, with\, tag\, t\, in\, P}{(C)\, Number\, of\, question\, in\, P}
\]
A candidate tag t is significantly relevant to ‘DevOps’ if its µ and v are higher or equal to certain thresholds. In our study we evaluate different range of values for significance and relevance where µ= \{0.005, 0.008, 0.010,  0.020, 0.030, 0.050\} and v= \{0.005, 0.010, 0.015, 0.020, 0.025, 0.030\}. We collected tag sets for 30 combinations of relevance and significance thresholds so that every tag set contains at least two tags (for significance 0.02, 0.03, 0.05, there were less than two tags with some relevance thresholds). Out of these 30 candidate tag sets, maximum 42 tags produced maximum 381,675 posts with the minimum threshold of significance (0.005) and relevance (0.005). The authors inspected each tag set manually to identify the tags which are most relevant with DevOps and discarded the tag sets which included tags not specific to DevOps. For example, the largest tag set of 42 tags contained tags bitbucket, sonarqube, digital-ocean, artifactory etc. which can be generic tools or applications not specifically related with DevOps. 

The target of manual analysis was to select a tag-set which has the highest number of DevOps related tags. With this aim, first two authors started reviewing all the tag-sets (TS1, TS2,…, TS30) together, starting from the tag-set with highest number of tags. Among the thirty selected tag-sets, the top 5 sets TS1, TS2, TS3, TS4, TS5, TS6 had 42, 30, 28, 24, and 19 tags respectively. First two authors jointly reviewed each tag-set to identify any non-DevOps related tag and recorded such tags against each set. For example, if we denote non DevOps related tags of tag-set-N as XTS-N, then first two authors identified following non-DevOps related tags in each tag-set XTS1 = {bitbucket, sonarqube, digital-ocean, amazon-ecs }, XTS2 = {amazon-ecs}, XTS3 = {sonarqube, cloud}, XTS4 = {amazon-ecs}. For the tag-set TS5, all tags were found to be DevOps related and XTS5 = {}. Then all four authors reviewed the excluded tags from the top four tag-sets XTS1, XTS2, XTS3, XTS4 and after a deliberation session together agreed that those tag-sets can be discarded and TS5 can be the final tag-set since it had the highest number of tags without any non-relevant tag. This joint review in two steps were made possible because the authors started reviewing tag-sets from the largest set in descending order and hence, after reviewing five sets, a final tag-set was decided.

The final tag set \( \delta_{devops} \) with significance, µ=0.008 and relevance, v=0.010, contained following 19 tags related with DevOps:

\begin{tcolorbox}[flushleft upper,boxrule=1pt,arc=0pt,left=0pt,right=0pt,top=0pt,bottom=0pt,colback=white,after=\ignorespacesafterend\par\noindent]
\noindent\textbf{Tag set \( \delta_{devops} \)} = \{ansible, azure-devops, devops, Jenkins, kubernetes, terraform, chef, continuous-integration, ibm-cloud, gitlab, jenkins-pipeline, gitlab-ci, puppet, amazon-cloudformation, pipeline, azure-pipelines, jenkins-plugins, continuous-deployment, devops-services\} 
\end{tcolorbox}

\subsubsection{\textbf{Extract DevOps Posts}}

After generating the DevOps tag set \( \delta_{devops} \), we extract SO posts whose tag set contain a tag in \( \delta_{devops} \). Our built dataset from SO includes 174,693 DevOps related question posts and their metadata. 

\subsection{RQ$_1$ DevOps Topics in Developer Discussion}

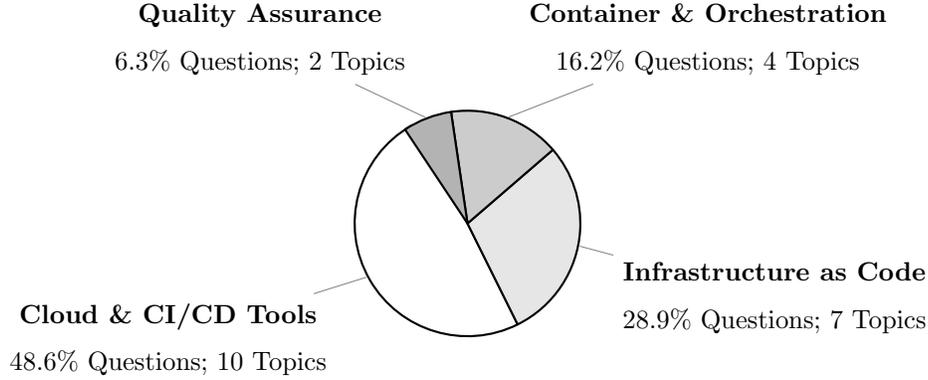
\begin{figure}[t]
	\centering
	\begin{tikzpicture}[scale=0.5]-
    \pie[
        /tikz/every pin/.style={align=center},
        text=pin, number in legend,
        explode=0.0,rotate=120,
        color={black!0, black!10, black!20, black!30,black!40},
        ]
        {
            49/\bf{Cloud \& CI/CD Tools}\\48.6\% Questions; 10 Topics,
            29/\bf{Infrastructure as Code}\\28.9\% Questions; 7 Topics,
            16/\bf{Container \& Orchestration}\\16.2\% Questions; 4 Topics,
            7/\bf{Quality Assurance}\\6.3\% Questions; 2 Topics
        }
    \end{tikzpicture}
	\caption{\% of questions and \# of topics per topic category}
	\label{fig:distribution_of_questions_topic_pie_chart}
\end{figure}

\begin{figure}
\includegraphics[scale=0.7]{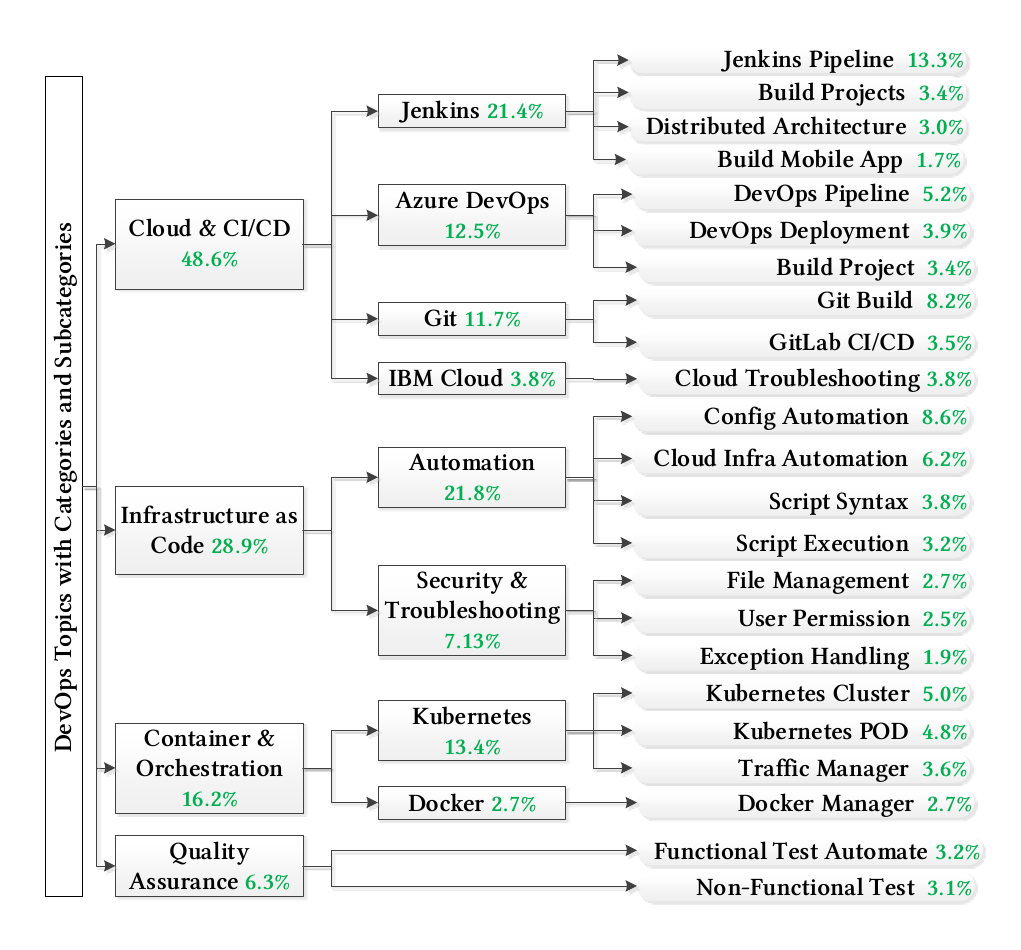}
\caption{DevOps topics categories and sub-categories}
\label{fig:DevOps_Topics}
\end{figure}

\subsubsection{\bf{Approach:}} Topic modeling has been increasingly used in empirical studies to mine unstructured data \cite{mens2014evolving}; previous works show that topic modeling can generate meaningful topics in software engineering research by mining textual documents \cite{Chen-SurveyTopicInSE-EMSE2016} \cite{Sun-ExploreTopicModelSurvey-SNPD2016}. Now we elaborate the process of topic modeling.

\nd{\ul{Preprocess  DevOps Post.}} We remove irrelevant information and noise from our dataset before applying topics modeling by following noise reduction steps adopted in previous works \cite{bagherzadeh2019going, Barua-2012, Bowen-2017}. We focus only on the post title for removing the noise from our dataset and the preprocessing steps are as follows. First, we remove the code snippets, HTML tags such as <p></p>, <html></html>, url tags, punctuation, and then stop words (a, an, is, the), numeric and non-alphabetic character by using NLTK stopwords corpus \cite{website:nltk}. Besides, bigram models were developed by using Gensim~\cite{Radim-gensim-LREC2010} since bigram models enhance the quality of text processing as reported by Tan et al~\cite{Talbigrams-2002}.  Moreover, we use Porter stemming to reduce words to their stemmed representation. For example, “configuration”, “configure”, and “configured” all reduce to “config” \cite{porter-steming-handbookchapter-1997}.

\nd{\ul{Model and Label DevOps Topic.}}
We formulate the topics using Latent Dirichlet Allocation (LDA)~\cite{Blei-LDA-JournalMachineLearning2003}. We used MALLET~\cite{McCallum-MALLET-Web2002} with Latent Dirichlet Assignment (LDA) Topic Modeling to group our DevOps posts into topics. LDA is intended to determine the optimal K-Topic number, whereby LDA generates detailed topics that are difficult to analyze if the topics are too large, but generates overlapping topics if the topics are too small.
We ran experiments with a broad range of values K = \{5, 10, 15, 20, 25, 30, 35, 40, 45, 50 \}, iteration (I) = \{100, 500, 1000, 1500 \} and we also set the range of the hyper-parameter values alpha = \{0.05, 0.1, 0.5, 1, 5, 10, 50\}/K, beta = \{0.01, 0.05 \} and observed the coherence value for each run following the configuration options provided in \cite{Bagherzadeh-BigdataTopic-FSE2019} \cite{Briggers-ConfigureLDA-EMSE2014}. The highest coherence value is \{0.4676 \} found for 30 topics, iteration 1000 and the value of the hyper-parameters alpha 5 and beta 0.05. With these parameters, the value of K is decided as 30. 

\nd{\ul{Generate the DevOps Topics.}} We found total 30 set of keywords from the SO dataset, each keyword set is candidate for a topic label. Each of the first two authors separately checked the 10 most popular keywords for each topic and checked at least 20 posts with highest correlation values with the topic. The other two authors reviewed the labels. Following previous work~\cite{Hudson-CardSorting-InteractionDesignFoundation2013,opencardsort-Fincher,Barua-StackoverflowTopics-ESE2012}, we use the open card sorting method~\cite{Hudson-CardSorting-InteractionDesignFoundation2013,opencardsort-Fincher} to mark the topic names. After agreement on the labeling, we merged a number of topics since their keywords had synonymous meanings. For example, two separate topics had top keyword set Topic15=\{git, repository, gitlab, jenkin, push, clone, remote, branch, commit, file\} and Topic12=\{jenkin, build, branch, trigger, pipeline, git, gitlab, commit, github, merge\}. But they are essentially similar topics related with Git merging and building using Jenkins or GitLab, and so they were merged into a single topic labeled Git Build. In the end, we found 23 different topics for our dataset. 
After confirming the 23 topics, we revisited all topics to categorize into higher levels. 

\nd\ul{Tool names in topic labels.} While labeling the topics and category names, we initially considered to give tool-agnostic name, e.g., instead of creating a sub-category of Jenkins, the tool-agnostic name could be Continuous-Integration. But after careful review of the top hundreds high correlation valued posts in those categories, it was evident that the topics did not discuss about generic continuous integration of other tools, rather the discussion was specific to the Jenkins tools and related pipeline or build processes. So, finally, we left the topic and sub-category names to be tool specific whenever was relevant. As mentioned in Section \ref{sec:related-work}, there are other studies related with similar topic modeling of continuous software engineering processes which attempted to avoid tool specific topics, but ended up having tools names in topics and categories \cite{Mansooreh-EASE2020}.

\subsubsection{\bf{Results}} 
We grouped the DevOps related questions into 23 unique topics and grouped the topics into four high-level categories: Cloud \& CI/CD Tools, Infrastructure as Code, Container \& Orchestration, and Quality Assurance. From Figure \ref{fig:distribution_of_questions_topic_pie_chart} shows the distribution of questions and topics under four major categories. Among the categories, Cloud \& CI/CD Tool that has the highest percentage of questions (48.6\%) and the highest number of topics (10 topics), followed by Infrastructure as Code (28.9\% of the questions, 7 topics), Container \& Orchestration (16.2\% questions, 4 topics), and Quality Assurance (6.3\% questions, 2 topics). 

Figure \ref{fig:DevOps_Topics} shows the 23 DevOps topics under four major categories ordered based on the distribution of questions (e.g., Cloud \& CI/CD tools is the top and Quality Assurance is the bottom category). There are sub-categories under each major category (except Quality Assurance), e.g., Infrastructure as Code has sub-categories of Automation and Security \& Troubleshooting. The final topics are placed under those sub-categories, e.g., Automation sub-category have four topics. Now we discuss the DevOps topics under their categories.

\nd\bf{$\bullet$ Cloud \& CI/CD Tools} In this category (48.6\% questions) DevOps engineers discuss about deployment, continuous integration/continuous delivery (CI / CD), Jenkins workflow and build, test and debug project with Jenkins for on-premises and cloud project, etc. This category contains ten topics and these topics are grouped into four sub-categories: Jenkins, Azure, Git and IBM.

\begin{inparaenum}[(i)]
\item \bf{{Jenkins (21.4\% questions)}} This sub-category is divided into four topics areas: 
\begin{inparaenum}[(1)]
\item \it{Jenkins Pipeline (13.3\% question)} contains discussion on events or tasks that are linked in a sequence to do quick software release and create codes that automatically implement new software versions  (e.g. (\dq{40454558})). 
\item \it{Build Mobile App (1.7\%)} contains questions about continuous build, test and debugging of the mobile applications with Jenkins tools. (e.g. (\dq{19426477})) 
\item \it{Build Projects (3.4\%)} contains discussion about end to end project building using Jenkins, build fail and error raised in Jenkins, etc. (e.g. (\dq{34130908})). 
\item \it{Distributed Architecture (3.0\%)} discusses about installation and build process troubleshooting for distributed master slave nodes (e.g. (\dq{45845662})).
\end{inparaenum}

    \item \bf{{Azure DevOps (12.5\% questions)}} sub-category contains discussions about Microsoft's DevOps and CI/CD tools. It covers three topics:
    \begin{inparaenum}[(1)]
    \item \it{Build Project (3.4\%)} 
    discusses about Azure DevOps automatic build, release, debug, troubleshooting questions (e.g. (\dq{62041467})).
    \item \it{DevOps Deployment (3.9\%)} contains questions about Azure DevOps web/app service deployment, deployment configuration, and app deployment difficulties (e.g. (\dq{28638945})).
    \item \it{DevOps Pipeline (5.2\%)}
    contains discussion about Azure DevOps project setup of an entire CI/CD pipeline (e.g. (\dq{58734429}))
    \end{inparaenum}

\item \bf{{Git (11.7\% questions)}} sub-category contains two topics about the popular source code management tool and associated integration tool:
\begin{inparaenum}[(1)]
\item \it {Git build (8.2 \% of the questions)} topic contains discussions about git configuration, build project code, GitHub project difficulties, tracking changes in the source code, (e.g. (\dq{58473877}))
\item \it {GitLab CI/CD (3.5 \% of the questions)} contains questions regarding GitLab setup for build projects, CI/CD, and continuous testing as an integration tool (e.g. (\dq{57340112})).
\end{inparaenum}

\item \bf{{IBM Cloud (3.8\% questions)}} sub-category covers only one topic:
    \begin{inparaenum}[(1)]
    \item \it {Cloud Troubleshooting (3.8\%)} contains questions about build, deployment, security management, in IBM's cloud platform for DevOps (e.g. (\dq{45775107})).
    \end{inparaenum}    

\end{inparaenum}

\nd\bf{$\bullet$ Infrastructure as Code} This is the second highest category which covers 28.9\% of the questions. It contains discussion about server configuration, cluster configuration and difficulties, file management, user access policy, infrastructure configuration scripts and troubleshooting, etc. This category has seven topics under two sub-categories as discussed below. 

\begin{inparaenum}[(i)]
    \item \bf{{Automation (21.8\% questions)}} Sub-category contains four topics:
    \begin{inparaenum}[(1)]
        \item \it{Configuration Automation (8.6\%)} discusses about on-premise infrastructure automatic configuration (i.e. make ansible-playbook, run the playbook for the given host etc. (\dq{45354448})).
        \item \it{Cloud Infrastructure Automation (6.2\%)} contains questions regarding automatic cloud configuration including hardware, software, networking components, operating system (OS), and data storage components using different automation tools such as Terraform, Cloudformation etc.
         \item \it{Script Syntax (3.8\%)} contains common troubleshooting questions related with different scripting issues (\dq{28963585}).
        \item \it{Script Execution (3.2\%)} is related with scripting, but the topic discusses more about infrastructure script execution procedures and challenges (e.g., permission settings, access controls of network devices or containers from certain scripts etc.). 
    \end{inparaenum}
    
    \item \bf{{Security \& Troubleshooting (7.13\% questions)}} contains discussion about security across the DevOps processes. This category contains three traditional IT infrastructural related topics whose titles are self-explanatory: File Management (2.7\% posts, e.g. \dq{59468464}), Exception Handling (1.9\% posts, e.g., \dq{53349416}) and User Permission (2.5\%, e.g., \dq{39473843}). 
    
\end{inparaenum} 

\nd\bf{$\bullet$ Container \& Orchestration} contains 16.2\% of DevOps posts that discuss about automating the deployment, scaling, networking, and management of containers. This category comprises of two sub-categories.

\begin{inparaenum}[(i)]
    \item \bf{{Docker (2.7\% questions)}} only one topic is covered by this sub-category:
        \begin{inparaenum}[(1)]
    \item \it{Docker Manager (2.7\%)} provides the ability to write code in one version and ship it to any platform, it provides the complete containerization.
    \end{inparaenum}
    
    \item \bf{{Kubernetes (13.4\% questions)}} is an open-source system for orchestration of containerized applications. There are three topics under this sub-category:
    \begin{inparaenum}[(1)]
    \item \it{Kubernetes Cluster (5.0\%)}
    is a collection of nodes that run containerized applications. 
    \item \it{Kubernetes POD (4.8\%)} discusses about Kubernetes cluster control management, Kubernetes cluster metric monitoring, operation (e.g. (\dq{52066896})).
    \item \it{Traffic Management (3.6\%)}
    contains discussion about resilience in Kubernetes with advanced traffic management, traffic routing, traffic shaping, and traffic monitoring (e.g. (\dq{60964361})).
    \end{inparaenum}
\end{inparaenum} 

\nd\bf{$\bullet$ Quality Assurance} contains 6.3\% of the questions. This category describes various aspects of testing in DevOps environment including automatic testing configuration, performance testing, unit and other testing issues. It contains two topics:

 \begin{inparaenum}[(1)]
    \item \it{Functional Test Automation (3.2\%)} topic contains posts on continuous integration testing, unit testing configuration, acceptance testing, setting up automatic functional testing and issues (e.g., Running iOS project unit tests from Jenkins does not produce any output (\dq{10751086}), dotMemory Unit Standalone launcher with XUnit hang at the end of the test (\dq{53535857})).
    \item \it{Non-functional Test Automation (3.1\%)}
    contains discussion about load test, performance test, Disaster Recovery, Security, Accessibility, Usability and Operational testing (e.g., How to publish Rough Auditing Tool for Security (RATS) xml results in Jenkins (\dq{14585222}))
    \end{inparaenum}
    
\begin{tcolorbox}[flushleft upper,boxrule=1pt,arc=0pt,left=0pt,right=0pt,top=0pt,bottom=0pt,colback=white,after=\ignorespacesafterend\par\noindent]
\noindent\textbf{RQ$_1$ What types of topics are discussed about DevOps in SO?} We found 23 topics that are divided into four categories: Cloud \& CI/CD Tool, Infrastructure as Code, Container \& Orchestration, and Quality Assurance. Cloud \& CI/CD Tools category covers nearly 49\% questions. Jenkins Pipeline under Cloud \& CI/CD Tools category is the most discussed topic with 13.3\% of total posts, followed by Configuration Automation topic (8.6\% posts) under Infrastructure as Code category. 
\end{tcolorbox}

\subsection{RQ$_2$ DevOps Topic Popularity vs Difficulty}\label{sec:rq-topic-difficulty}

\begin{table*}
  \centering
   \caption{DevOps topics, their popularity, and difficulty}
  \resizebox{\columnwidth}{!}
  {%
    \begin{tabular}{lrrrrrrr}
    \toprule{}
    \multirow{3}{*}{\textbf{Topic}} & 
     \multicolumn{3}{c}{\textbf{Popularity score}} & \multicolumn{2}{c}{\textbf{Difficulty score}} \\
    {}  & 
    \textbf{Avg} & \textbf{Avg} & \textbf{Avg}  & \textbf{W/O acc.} & \textbf{Med. Hrs} \\
    {}  & 
    \textbf{view} & \textbf{fav.} & \textbf{score}  & \textbf{ans.} & \textbf{acc.} \\

    \hline
    \multicolumn{6}{l}{\textbf{Cloud \& CI/CD Tool}} \\
    \hline
    Jenkins Pipeline &	2673 &	0.53 &	2.19 & 61.7\% &	194 \\
    Jenkins Distributed Architecture &	2332 &	0.52 &	1.90 & 64.7\% &	174 \\
    Git build   &	2402 &	0.71 &	2.52 & 60.4\% &	229 \\
    Jenkins Build Projects &	2035 &	0.46 &	1.78 & 63.3\% &	154 \\
    Jenkins Build Mobile App &	1543 &	0.74 &	2.38 & 61.1\% &	248 \\
    GitLab &	1416 &	0.57 &	2.09 & 60.1\% &	250 \\
    Azure Build Project  &	1391 &	0.46 &	1.93 & 44.4\% &	164 \\
    Azure DevOps Deployment &	933 &	0.69 &	1.75 & 56.8\% &	142 \\
    Azure DevOps Pipeline  & 607 &	0.24 &	1.09 & 45.8\% &	132 \\
    IBM Cloud Troubleshooting & 421 &	0.16 &	0.66 & 59.5\% &	74 \\

    \hline
    \multicolumn{6}{l}{\textbf{Infrastructure as Code}} \\
    \hline 
    File Management &	2805 &	0.40 &	2.05 & 58.1\% &	191 \\
    Script Execution &	2603 &	0.39 &	1.75 &	59.1\% & 135 \\
   User Permission  &	2509 &	0.63 &	2.23 & 58.6\% &	216 \\
   Configuration Automation & 2125 &	0.45 &	1.87 & 51.6\% &	145 \\
    Exception Handling  &	1970 &	0.42 &	1.74 & 65.6\% &	132 \\
    Script Syntax  &	2018 &	0.29 &	1.44 & 44.9\% &	73 \\
   Cloud Infrastructure Automation  &	1054 &	0.38 &	1.62 & 57.4\% &	133 \\
\hline
    \multicolumn{6}{l}{\textbf{Quality Assurance}} \\
    \hline
    Non-func Test Automation &	2062 &	0.46 &	1.95 & 63.7\% &	251 \\
    Func Test Automation  &	1400 &	0.50 &	1.65 & 63.9\% &	197 \\

    \hline
    \multicolumn{6}{l}{\textbf{Container and Orchestration}} \\
    \hline
    Docker Manager &	1435 &	0.60 &	1.94 & 59.1\% &	138 \\
    Kubernetes POD  &	1381 &	0.46 &	1.89 & 57.2\% &	127 \\
    Kubernetes Cluster  & 1020 &	0.42 &	1.44 & 59.9\% &	106 \\
    Traffic Management &	922 &	0.38 &	1.38 & 61.8\% &	107 \\
    \hline
    \textbf{Average}  & \textbf{1695} & \textbf{0.47} & \textbf{1.8} & \textbf{58\%} & \textbf{161.4}  &\textbf{} &  \\
    \bottomrule
    \end{tabular}%
    }
  \label{tab:topicPopularityDifficulty}%
\end{table*}%

\subsubsection{Approach}
We compute three metrics to measure popularity of each topic: \begin{inparaenum} \item Average number of views for all posts of a topic, \item Average number of questions of a topic marked as users' favorite, and \item Average score of questions of a topic.
\end{inparaenum}
We determine the difficulty of the topic using two metrics of the question, \begin{inparaenum}
\item Percentage of the questions without any accepted answer, and \item Average time needed to receive an accepted answer.   
\end{inparaenum} These mentioned five metrics were also previously used in several research papers to compute the popularity and difficulty of topics found in SO posts \cite{Rosen-MobileDeveloperSO-ESE2015} \cite{Bagherzadeh-BigdataTopic-FSE2019} \cite{abdellatif2020challenges}. Finally, we also assess the correlation between each of three topic popularity and two difficulty
metrics. We use Kendall’s \( \tau\) correlation measure \cite{kendall1938new}. Unlike Mann-Whitney
correlation \cite{Kruskal-Wilcoxon-JASS1957}, Kendall’s \( \tau\) is not susceptible to outliers in the data. 

\subsubsection{Results}
\nd\textbf{\ul{Topic Popularity,}}
Table \ref{tab:topicPopularityDifficulty} shows three popularity metrics for each DevOps topic: average number of \begin{inparaenum}
   \item view count \item favorite count \item score.
\end{inparaenum} Topics are ranked in descending order by average view count under each category.

File Management under Infrastructure as Code category has the highest number of average view count (2,805). This topic is comprised of different file management issues which can vary from general purpose file to various tool specific files such as Dockerfile, build log file, timestamped war file e.g. \dq{43161244}. Second most viewed topic is Jenkins Pipeline under Cloud \& CI/CD Tools category (2,673 views). This is a merged topic comprising of scripted and descriptive Jenkins pipeline, overall Jenkins job flow, and environment variable settings. Example posts are \dq{56887022}, \dq{46429657}. 

The topics Script Execution (view 2,603) and User Permission (view 2,509) from Infrastructure as Code category are the third and fourth most viewed topics respectively. Both of these topics are related with general infrastructure problems related with any DevOps tools e.g., "How to call batch/groovy script located in jenkins slave from "Execute system Groovy script"?" (\dq{15172048}), "How to make Windows START command accept standard input via pipe and pass it on to the command it invokes?" (\dq{34512012}). 
Git Build has the highest average favorite posts (2.52) followed by Build Mobile App (favorite score 2.38). IBM Cloud troubleshooting is the least viewed topic (421 views) followed by Azure DevOps Pipeline (607 views).  

\nd\textbf{\ul{Topic Difficulty.}} Table \ref{tab:topicPopularityDifficulty} contains the difficulty metrics per topic \begin{inparaenum}
   \item Percentages of question without any accepted answer, and 
   \item Median hours taken to receive an accepted answer.
\end{inparaenum}

Among all the topics, Exception handling from infrastructure as Code category is ranked the most difficult topic because around 65.6\% of its questions remain without accepted answers, while those that get an accepted answer normally had to wait a median of 132 hours. The following most difficult topic is Jenkins Distributed Architecture whose unaccepted post rate is 64.7\% with a median wait time of 174 hours for accepted answers. Non-functional Test Automation topic has the highest median waiting time of 251 hours for accepted answer. A non-functional test automation question asked 4 years ago still got no accepted answer \dq{44236838}, whereas another question with 20K views (\dq{18779450}) from the same topic got its accepted answer after four years of asking. Questions from GitLab topic have median waiting time for 250 hours for accepted answer. For example, a question \dq{31687499} was asked more than 6 years ago, still did not get any accepted answer.

\nd\textbf{\ul{Correlation between Topic Popularity and Difficulty.}} Our results from topic popularity and difficulty does not indicate any positive or inverse relationship between all the popularity and difficulty metrics of a topic. For example, the most popular (viewed) topic File Management has 58\% unsolved questions. Similarly, the least popular topic of Cloud Troubleshooting-IBM also has 59\% unsolved questions. 
On the contrary, the most difficult topic Exception Handling (65.6\% unaccepted questions) has an average view of 1970 and favorite score 0.42, whereas Configuration Automation topic has similar view and favorite score (2125 view and 0.45 score), but it is the fourth easiest topic with 51.6\% unaccepted questions. Popularity does not correlate with solution of questions.

Similar statistics can be found from Table \ref{tab:correlation} which shows six correlation measures (using Kendall’s \( \tau\)) between topic difficulty and popularity in Table \ref{tab:topicPopularityDifficulty}. All six correlation coefficients are positive and three of them (related with median hours to acceptable answers and popularity metrics) are statistically significant with more than 95\% confidence level. From this correlation, it shows that more popular topics can take more time to get an acceptable answer. But there is no significant correlation between popularity and percentage of questions without accepted answer.

\begin{table}[t]
  \centering
  \caption{Correlation between topic popularity \& difficulty}
    \begin{tabular}{lrrr}\toprule
    Coefficient/p-value & \bf{View} & \bf{Favorites} & \bf{Score}\\ \midrule
    \bf{\% w/o acc. ans.} &   0.138/0.373    & 0.173/0.254      &  0.123/0.413 \\
    \bf{Med hrs to acc. ans.} &    0.375/0.012   &   0.566/0.0002     & 0.655/1e-05  \\
    \bottomrule
    \end{tabular}%
  \label{tab:correlation}%
\end{table}

\begin{tcolorbox}[flushleft upper,boxrule=1pt,arc=0pt,left=0pt,right=0pt,top=0pt,bottom=0pt,colback=white,after=\ignorespacesafterend\par\noindent]
\noindent\textbf{RQ$_2$ How do the popularity and difficulty of DevOps topics vary?} 
File Management topic, from the infrastructure as Code category, is the most popular in terms of post views. Exception Handling, from the same category, is the most difficult in terms of getting an accepted answer. Non-functional Test Automation takes the longest 251 hours to get an acceptable answer. There was no statistically significant inverse correlation between popularity and difficulty of topics. 
\end{tcolorbox}

\section{Survey of DevOps Practitioners}\label{sec:survey}
In this section, we report a survey of DevOps professionals that we conducted to validate our empirical study findings and to get additional insights about DevOps challenges, if there are any. By analyzing the survey responses, we answer two research questions:
\begin{enumerate}[label=\bf{RQ\arabic{*}.}, start=3]
\item How the industry practitioners agree with our empirical study findings?
\item What additional insights about DevOps challenges can we get from open-ended responses of survey participants?
\end{enumerate}
\nd\bf{{Survey Participants.}} To conduct the survey, we targeted industry professionals in small, medium, and large corporations. To select the participants, we combined judgemental sampling \cite{vogt2015sage} with snowball sampling \cite{creswell2017research}. In judgemental sampling, we start with participants whom we knew had already experience in DevOps as architect, manager, developer, or operational engineer. 

Around 55\% of participants had more than 10 years of professional experience, and 95\% had at least two years experience. Hence, the participants in general had sufficient hands on professional experience. They had diversified roles in their organizations: 35\% solution architects and 15\% managers (delivery or operations), 30\% were development engineers, and 15\% were operations engineers.
Around 48\% of participants came from small organizations (1-50 employees), 32\% from small to medium organizations (51-150 employees), and the rest (20\%) from medium to large organizations (150+ employees). 45\% of participants evaluated their organization to be in the mid-level in terms of DevOps maturity and 50\% consider they have advanced implementation and practice of DevOps. Distribution of both the size and maturity of the organizations are not skewed and can be considered well-balanced representation of general organizations. The survey questions and anonymized responses are shared in our replication package \cite{replciation-package}.   

\nd\bf{{Survey Questions.}} All survey questions are listed in Table \ref{tab:survey-questions-1} and \ref{tab:survey-questions-2}. Table \ref{tab:survey-questions-1} contains the fixed answer questions (multiple choice and checkboxes) and Table \ref{tab:survey-questions-2} contains open-ended questions. The detailed motivation of each question is discussed in the following subsections along with the results and analysis.
\begin{table}[t]
  \centering
  {\caption{Survey questions with fixed answers. Target research questions are denoted in the RQ column.\label{tab:survey-questions-1}}}
  
  \resizebox{\textwidth}{!}{%
    \begin{tabular}{p{.4cm}p{.6cm}p{9.3cm}p{2cm}}\toprule
    \# & \bf{RQ} & \bf{Survey Question} & \bf{Ans. Type}\\ \midrule
Q1 & Info & What is the size of your engineering organization? & Multiple Choice \\ 
Q2 & Info & On a scale of 1 to 5 where your team or organization positions in practice of DevOps ? & Multiple Choice \\ 
Q3 & Info & What is the earliest year your organization started devOps practices? & Multiple Choice \\ 
Q5 & RQ1 & What tools do you use as part of your devOps practice? (You can write additional tools in other box as comma separated) & Checkboxes \\ 
Q6 & RQ4 & What do you think the most critical part of devOps life cycle? & Multiple Choice \\ 
Q8 & RQ4 & When a new project is ready to go live, which devOps part you initiate first? & Multiple Choice \\ 
Q9 & RQ2 & Please choose difficulty level for each of the tools/topics below: \newline
[Build (Ant, Maven, Gradle)]  \newline
[Quality Assurance (SonarQube, Selenium, JMeter, ...)] \newline
[Integration Pipeline (Jenkins, ...)] \newline
[Configuration Automation (Ansible, Chef, Puppet, ...)] \newline
[Containerization \& Orchestration (Docker, DockerSwarm, Kubernetes, ...)] \newline
[Cloud Infra Automation (Terraform, CloudFormation, ...)] \newline
[Monitoring (Splunk, ELK, Grafana/Prometheus), ...]  & Multiple Choice Grid \\ 
Q12 & RQ2 & How often your developers face problems related with basic IT Infra (e.g., repository or file settings, user permission settings, scripting syntax or execution permission etc.) & Multiple Choice \\ 
Q13 & RQ2 & How often you get solutions from community or Stack Overflow about the devOps problems you  face? & Multiple Choice \\ 
Q14 & RQ1 & Which cloud services do you use & Checkboxes \\ 

    \bottomrule
    \end{tabular}%
    }
\end{table}
\begin{table}[t]
  \centering
  {\caption{Survey Questions with open ended answers. Most of these were related with RQ4 (industry insights) and some questions also helped analyze RQ1 as denoted under column RQ\label{tab:survey-questions-2}}}
    \begin{tabular}{p{.4cm}p{1.5cm}p{8cm}p{2cm}}\toprule
    \# & \bf{RQ} & \bf{Question} & \bf{Type}\\ \midrule
Q4 & RQ4 & What organizational process/training you follow to adapt devOps practices and tools? & Open-ended answer \\ 
Q7 & RQ4+RQ1 & What difficulties do you face in this critical phase of devOps that you work on? & Open-ended answer \\ 
Q10 & RQ4+RQ1 & What are the advantages of the continuous integration tool you primarily use (e.g. Jenkins or Gitlab CI etc.)? & Open-ended answer \\ 
Q11 & RQ4+RQ1 & What challenges you face while using your primary integration tool (you may also mention status of community support)? & Open-ended answer \\ 
Q15 & RQ4 & Do you think Cloud infra automation (Terraform, CloudFormation etc.) will be more used in future? If so, why? If not, why? & Open-ended answer \\

    \bottomrule
    \end{tabular}%
  \vspace{-5mm}
\end{table}

\subsection{RQ$_3$ Validation of Empirical Study Findings}
\subsubsection{\textbf{Approach}} We designed the survey questions around the key findings of our first two research questions as follows.

\textbf{Questions regarding RQ1.} We attempted to match the concepts found in DevOps literature such as continuous software engineering processes (e.g., CI/CD, continuous testing etc.) with the industrial practice of the professionals - which processes they start first in DevOps initiative, and what are the crucial processes and their challenges in DevOps. Moving forward from the DevOps concepts, there were questions about the popular tools - the list, category, and usage of tools by the practitioners. We listed and grouped the tools based on the findings of the empirical study. In every case, we also added option to provide new tools or categories. 

\textbf{Questions regarding RQ2.} As we found from the empirical study that continuous integration tools are one of the most discussed topic, we asked about advantages and challenges of the integration tools using open ended questions.

\subsubsection{\textbf{Results}}

\begin{table*}
  \centering
  \rowcolors{3}{}{lightgray!30}
  \caption{Mapping of key findings between the empirical study and the survey responses}
  \resizebox{\textwidth}{!}{%
    \begin{tabular}{p{6cm}|p{6cm}|p{6cm}}\toprule
    \bf{Empirical Study Finding (E)}
    & \bf{Survey Finding (S)} 
    & \bf{Remarks (R)} \\\midrule
    \multicolumn{3}{c}{\bf{RQ1. What types of topics are discussed about DevOps in SO?}} \\
    
    E1. 64.8\% of questions in DevOps are related with specific tools (under Cloud \& CI/CD and Container \& Orchestration categories) and total 14 out of 23 i.e. 61\% topics were tool specific. Automation (with tools) is the most discussed sub-category (21.8\% of all DevOps questions). &
        S1 (Q5, Q7, Q14). 20 respondents listed total 39 tools and 9 cloud services they use related with DevOps. One participant responded \textit{"Too many tools and practices. Each team must explore do trial and error and settle on what best works for the team."} & 
        R1. There are plethora of useful tools for practicing DevOps, and proper usage of tools is a critical part for successful DevOps implementation. \\
    
    E2. The CI/CD engine Jenkins (21.4\% questions), the second most discussed sub-category, is the most discussed tool. Another CI/CD tool GitLab CI/CD had 3.5\% questions (one sixth of Jenkins) in SO. &
        S2(Q5) Jenkins is used by 65\%, GitLab CI/CD is used by 35\%, and Circle CI/CD is used by 5\% of respondents. &
        R2. Jenkins usage is 2 times, but Jenkins related questions in SO are 6 times of GitLab CI/CD. This could either mean GitLab CI/CD is a simple stable tool or Jenkins is more difficult to use tool. \\ 

    E3. Cloud platform topics: 12.5\% posts on Azure DevOps, 4.6\% Amazon CloudFormation 3.8\% IBM Cloud.  & 
        S3 (Q14)  Usage: Azure used by 30\%, Amazon CloudFormation used by 10\% participants. No participants use IBM Cloud. & 
        R4. Azure is used 3 times and also discussed in SO around 3 times of CloudFormation. The empirical and survey data seems to be congruent.  \\
    \midrule
    \multicolumn{3}{c}{\bf{RQ2. How do the popularity and difficulty of DevOps topics vary?}} \\
    
    E4. Jenkins is the most popular tool in terms of number of posts in SO. But Jenkins topics require higher range of waiting period to get acceptable answers (from 154 to 248 hours in average, whereas median waiting period is 150 hours). & 
        S4 (Q10, Q11). Most respondents mentioned Jenkins is easy to use. They also do not consider long answer time in SO to be a major problem. & 
        R4. From empirical study data, Jenkins appeared to be a more difficult topic (high unsolved questions and waiting time), but from practitioners' response, Jenkins rather seems to be a useful, easy to use tool.
        \\
    E5. Out of 174K DevOps posts in Stack Overflow, 101K posts (58\%) do not have solution and it takes 161.4 hours to get an accepted answer.  & 
        S5 (Q13). 68.4\% respondents perceive that Stack Overflow provides solutions \textit{most cases} and 10.5\% feels that they get solutions \textit{every time}. & 
        R5. The respondents seems to be getting more support from Stack Overflow than the empirical data presents. \\  
   
    \bottomrule
    \end{tabular}%
    }
  \label{tab:empiricalstudy_and_survey}%
\end{table*}%

In \tbl\ref{tab:empiricalstudy_and_survey}, we validate key findings from RQ1 and RQ2 against the results of RQ3 from the survey. We group the findings by the two RQs. For example, for RQ1, we have three consolidated key findings in \tbl\ref{tab:empiricalstudy_and_survey} (E1-E3), and we find three key insights (S1-S3) from our survey responses regarding E1-E3; and there are two findings compared from RQ2 (E4, E5). The left column in \tbl\ref{tab:empiricalstudy_and_survey} shows the key findings from the empirical study and the middle column shows the corresponding validated insights from the survey responses. For every finding pair, we provide our remark on the right column.

\begin{figure}[h]
\centering
\includegraphics[scale=0.70]{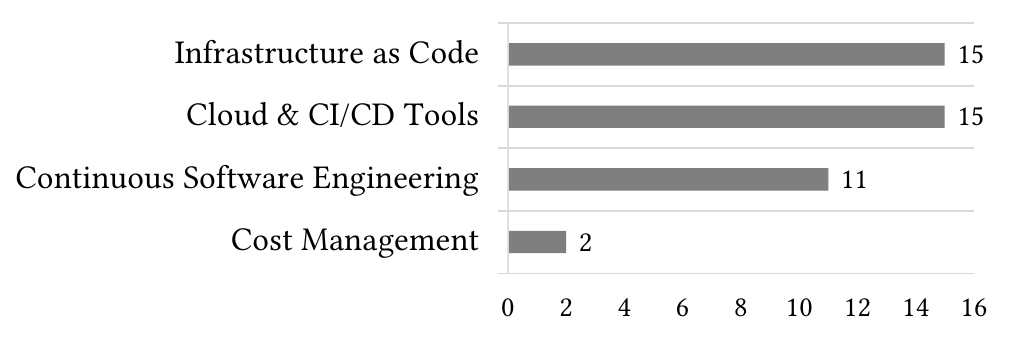} 
\caption{Responded topics in advantages of integration tool}
\label{fig:survey-q10}
\end{figure}

\begin{figure}[h]
\centering
\includegraphics[scale=0.7]{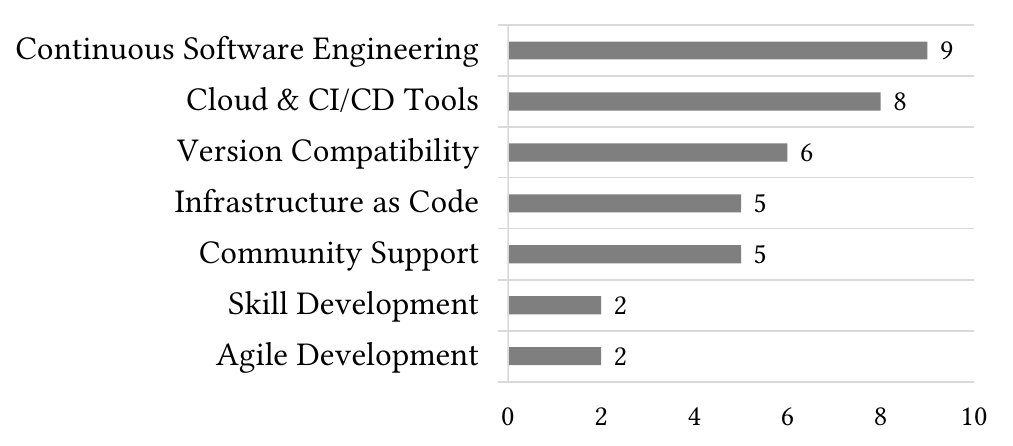} 
\caption{Responded topics in challenges of integration tool}
\label{fig:survey-q11}
\end{figure}

Among the overwhelming number of tools and posts regarding DevOps, in our quantitative study one tool particularly stands out which is Jenkins. Despite being the most popular tool, 61.1\% to 64.7\% of Jenkins related posts are unsolved in SO, and all four topics of Jenkins are included among the top ten most difficult topics of DevOps. With couple of open-ended questions (Q10, Q11), we wanted to validate whether practitioners also feel the similar difficulties in Jenkins. The topic analysis from those open-ended questions are shown in Figure \ref{fig:survey-q10}, and Figure \ref{fig:survey-q11}.

Slightly contrary to the research finding, the participants seemed to consider Jenkins a user-friendly helpful tool. The overall impression can be reflected by a response from a participant \textit{"It [Jenkins] is user-friendly, easy to install and does not require additional installations or components. It is free of cost. Easily Configurable. Jenkins can be easily modified and extended. It deploys code instantly, generates test reports. Platform Independent. Jenkins is available for all platforms and different operating systems, whether OS X, Windows or Linux. Rich Plugin ecosystem. The extensive pool of plugins makes Jenkins flexible and allows building, deploying and automating across various platforms."}

It is possible that because of the ubiquity of Jenkins, a lot of unsolved/unanswered questions are trivial or duplicate and not responded by the active community in SO. Consequently, we can conclude the ease of Jenkins, but we cannot conclude the reason of difference between the empirical study and survey. 

\begin{tcolorbox}[flushleft upper,boxrule=1pt,arc=0pt,left=0pt,right=0pt,top=0pt,bottom=0pt,colback=white,after=\ignorespacesafterend\par\noindent]
\noindent\textbf{RQ$_3$ How the industry practitioners agree with our empirical study findings?} The survey responses mostly align with the findings of RQ1 and RQ2. 61\% DevOps topics were specific tool related, and the participants also reported using 39  tools and 9 cloud services for DevOps. In terms of difficulty of the topics in SO, respondents seem to be more positive of solutions provided by SO. In SO analysis, we found Jenkins has 6 times posts of GitLab, but in survey, Jenkins was used 2 times of GitLab.
\end{tcolorbox}

\subsection{\textbf{RQ$_4$ More Insights about DevOps Challenges}} 

\subsubsection{\textbf{Approach}} While most of the survey questions were around the validation of the empirical findings, we also added five open-ended questions for qualitative analysis on challenges of DevOps processes (CSE topics), tools (CI/CD tools), and on future trend (cloud infrastructure automation) as listed in Table \ref{tab:survey-questions-2}. To analyze the responses, we applied an open coding approach \cite{miles_1994} that includes labeling of concerns/categories in textual contents. In our open coding, we followed the card sorting approach \cite{opencardsort-Fincher}. For example, the following response "it's error free and suitable for DR" has two different conceptual coherent quotes, "Error Free", and "Suitable for DR". The first quote refers to quality assurance, and second one 'disaster recovery' (DR) refers to automation of infrastructure as code in DevOps context. The first two authors coded the open-ended survey responses to extract potential categories independently. The level of agreement between the coders was calculated using Cohen's \( \kappa\) \cite{Cohen-Kappa-EducationalPsy1960}. The value of \(\kappa\) was 0.63 referring to a substantial agreement.

\subsubsection{\textbf{Results}} 
\begin{table}[h]
    \centering
    \caption{The topic or concern categories observed during the open-coding of the responses (\#TC = Total Categories, \#TP = Total number of participants, \#C = the number of code, \#P = the number of participants referring the code)}
    \resizebox{\columnwidth}{!}{%
    \begin{tabular}{r|r r|r r|r r|r r|r r|r r}
  
        \toprule
       \bf{Topic or Concern} & \multicolumn{2}{c}{Q04} & \multicolumn{2}{c}{Q07} & \multicolumn{2}{c}{Q10} & \multicolumn{2}{c}{Q11} & \multicolumn{2}{c}{Q15} & \multicolumn{2}{c}{} \\ 
       \cmidrule{2-11}
        \bf{Category} & \#C & \#P & \#C & \#P & \#C & \#P & \#C & \#P & \#C & \#P & \#TC & \#TP \\ \midrule
        
\bf{Agile Development} & 4 & 4 &  &  &  &  & 2 & 2 & 1 & 1 & 7 & 7 \\ \cmidrule{2-13}
\bf{Cloud \& CI/CD Tools} & 3 & 3 & 3 & 3 & 15 & 8 & 8 & 8 & 7 & 6 & 36 & 28 \\ \cmidrule{2-13}
\bf{Community Support} &  &  &  &  &  &  & 5 & 3 &  &  & 5 & 3 \\ \cmidrule{2-13}
\bf{Container \& Orch.} & 6 & 4 & 1 & 1 &  &  &  &  & 1 & 1 & 8 & 6 \\ \cmidrule{2-13}
\bf{Continuous SW Engg.} & 7 & 3 & 5 & 4 & 11 & 6 & 9 & 8 & 5 & 1 & 37 & 22 \\ \cmidrule{2-13}
\bf{Cost Management} &  &  &  &  & 2 & 2 &  &  & 1 & 1 & 3 & 3 \\ \cmidrule{2-13}
\bf{Infrastructure as Code} & 3 & 3 & 9 & 6 & 15 & 11 & 5 & 5 & 19 & 11 & 51 & 36 \\ \cmidrule{2-13}
\bf{Quality Assurance} & 1 & 1 &  &  &  &  &  &  & 2 & 2 & 3 & 3 \\ \cmidrule{2-13}
\bf{Skill Development} & 11 & 6 & 4 & 4 &  &  & 2 & 2 & 2 & 1 & 19 & 13 \\ \cmidrule{2-13}
\bf{Version Compatibility} &  &  &  &  &  &  & 6 & 3 &  &  & 6 & 3 \\ \midrule
\bf{Total} & 35 & 24 & 22 & 18 & 43 & 27 & 37 & 31 & 38 & 24 & 175 & 124 \\ \midrule

\bf{\# Categories/Question} & \multicolumn{2}{c}{7} & \multicolumn{2}{c}{5} & \multicolumn{2}{c}{4} & \multicolumn{2}{c}{7} & \multicolumn{2}{c}{8} & \multicolumn{2}{c}{} \\
        
        \bottomrule
    \end{tabular}%
    }
    \label{tab:open-coding}
\end{table}

\begin{figure}[h]
\includegraphics[scale=0.60]{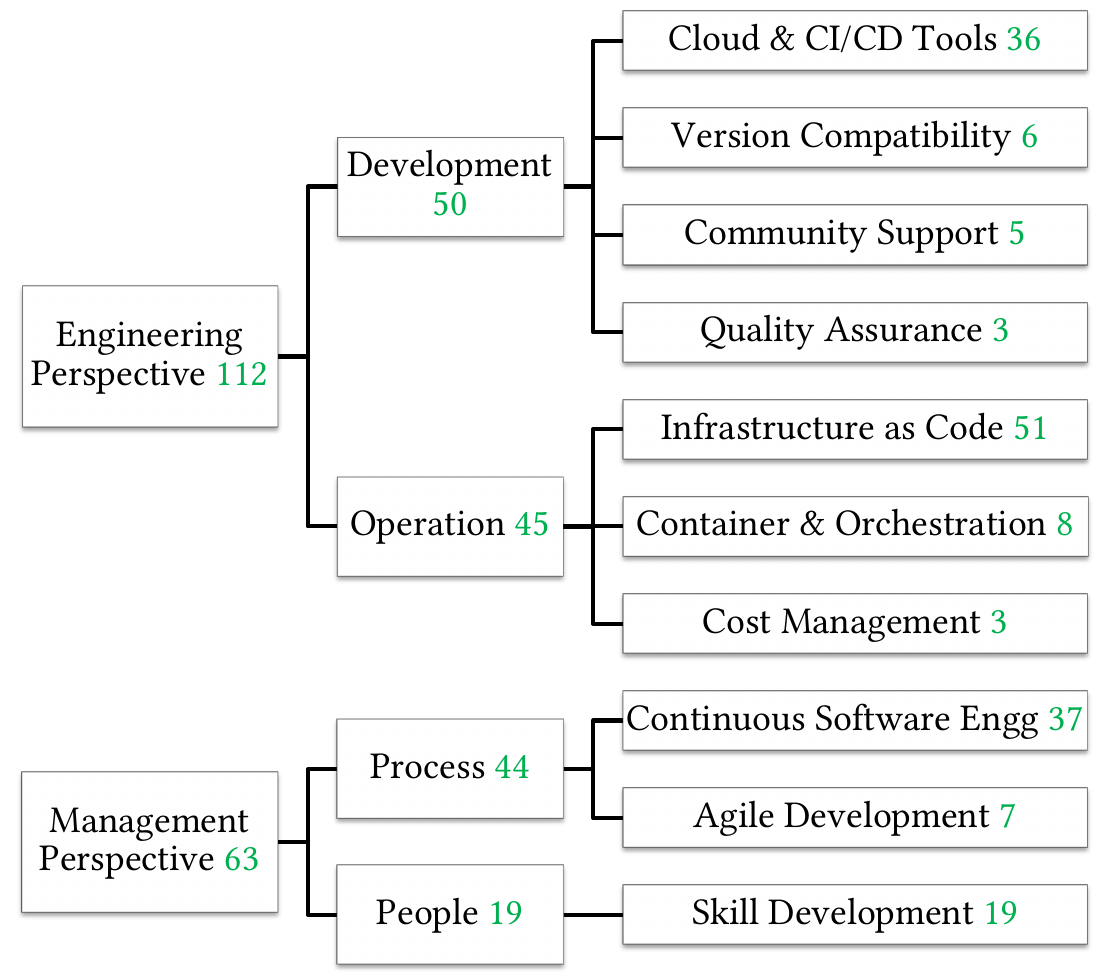} 
\caption{Hierarchy of topics observed in qualitative survey}
\label{fig:taxonomy-survey}
\end{figure}

The open coding of those survey question responses resulted in a total of ten topic or concern categories. Figure \ref{fig:taxonomy-survey} lists hierarchy and count of each topic found in the survey. We used conceptual map of \cite{Leite-DevopsSurvey-ACM2020} to categorize the topics. 
In Table \ref{tab:open-coding}, we show the number
of quotes for each topic as we found in our survey data. In following subsections, we explain the categories by each open-ended question (Q4 regarding organization DevOps training, Q7 about DevOps processes, Q10 about CI/CD tool advantages, Q11 regarding challenges of CI/CD tool, and Q15 about the opinion of cloud automation's future). 

\nd\textbf{Importance of Continuous DevOps Processes (Q6, Q8):} As noted in Background Section \ref{sec:related-work}, there are important continuous software engineering processes in the DevOps concept. The empirical study did not directly shed light on those processes. So we included few open ended questions in the survey to get industry feedback. 45\% of the respondents view continuous integration as most critical part of DevOps, whereas 25\% emphasized on continuous monitoring as most critical process. Others almost equally view continuous deployment, testing, and building. 

In a slightly different perspective, when asked which DevOps practices they start in any new project, the respondents were almost equally divided among all major processes of Continuous Testing (15\%), Containerization \& Orchestration (15\%), Configuration Automation (15\%), Complete Integration Pipeline (25\%). Here is an excerpt from a participant about the importance and risks of deployment as \textit{"So many things can go wrong during deployment. Deactivating the currently running version while deploying the new one is complicated with full ecosystem support. Without properly testing the deployment during staging, migration can go wrong. For large applications in Java, it still needs manual intervention."}

\begin{figure}[h]
\centering
\includegraphics[scale=0.70]{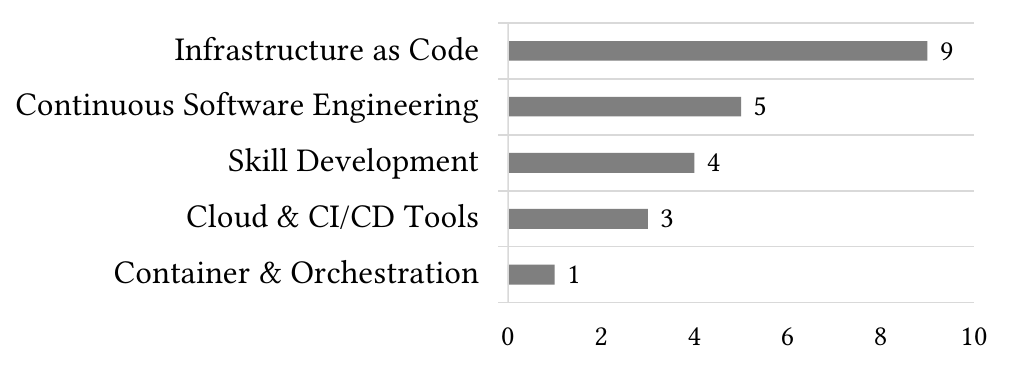} 
\caption{Continuous Software Engineering Topics}
\label{fig:survey-q7}
\end{figure}

\nd\textbf{Difficulties of DevOps Processes (Q7):} 30\% respondents mentioned the difficulties to continuous integration and testing among multiple cloud services (e.g., between AWS and Azure), and adaptation of various dependent and connected systems/tools is a major challenge. 30\% of practitioners think that keeping engineers aligned, and skilled with DevOps practices is challenging. The topics responded by participants is listed in Figure \ref{fig:survey-q7}.

\begin{table}[t]
  \centering
   \caption{Difficulty level of different tools set by DevOps continuous processes. Each colored bar denotes a difficulty level. Black = Not used, Green = Easy to use and learn, Magenta = Easy to use after some experience, Orange = Frequent troubleshooting needed, Red = Even SO cannot answer}
    \resizebox{\columnwidth}{!}{%
    \begin{tabular}{lr}
    \toprule{}
    \textbf{Process wise tools} & \textbf{Difficulty level}\\
    \midrule
    Build Tools	& \difficultyfivebars{42}{53}{0}{0}{5}		\\
    Integration and Pipeline Tools & \difficultyfivebars{26}{58}{11}{0}{5}	\\
    Containerization Tools	& \difficultyfivebars{0}{53}{32}{11}{5}	\\ 
    Monitoring Tools	& \difficultyfivebars{5}{42}{26}{11}{16} \\
    Quality Assurance Tools	& \difficultyfivebars{21}{47}{11}{5}{16} \\
    Configuration Automation Tools	& \difficultyfivebars{5}{32}{21}{5}{37} \\
    Cloud Infra Automation Tools & \difficultyfivebars{0}{32}{5}{11}{53}	\\
    \bottomrule
    \end{tabular}%
   }

  \label{tab:toolsDifficulty}%
\end{table}%

\nd\textbf{Difficulty Level of Process Specific Tools (Q9):} Since 60\% of DevOps topics were tool specific, instead of asking the difficulty level of the 23 general topics, we asked the practitioner their opinion on the tools they use. As also noted that there are numerous DevOps tools, we categorized the tools according to the DevOps processes and collected the industry feedback. Build tools found to be easiest (by 42\% respondents), while Containerization and Monitoring tools seemed to be most difficult (by 11\% participants), and Cloud infra automation tools were least used (53\% respondents did not use any such tool). The details are provided in the Table \ref{tab:toolsDifficulty}.

\begin{figure}[h]
\centering
\includegraphics[scale=0.7]{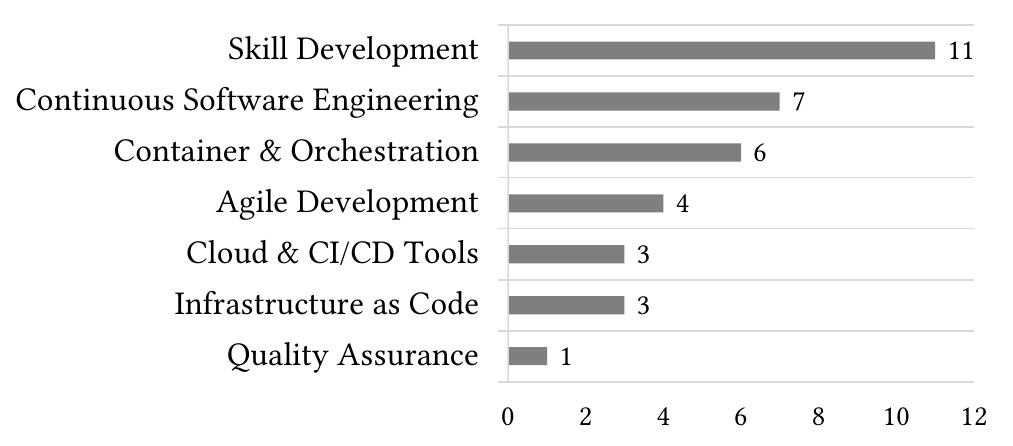} 
\caption{Skill Development Requirement Topics}
\label{fig:survey-q4}
\end{figure}

\nd\textbf{Skill Development in DevOps (Q4):} We wanted to know what organization are doing for training and skill development in DevOps area. 20\% practitioners think that lack of skill is the most common problem in DevOps. But only 5\% companies have formal training in DevOps. 10\% participants learn DevOps from online courses such as Udemy, Plural Sigh. One respondent mentioned \textit{"most of the knowledge transfer is done ad-hoc by the mentor or other senior engineers."}. Topics referred in the question is shown in Figure \ref{fig:survey-q4}.

\begin{figure}[h]
\centering
\includegraphics[scale=0.7]{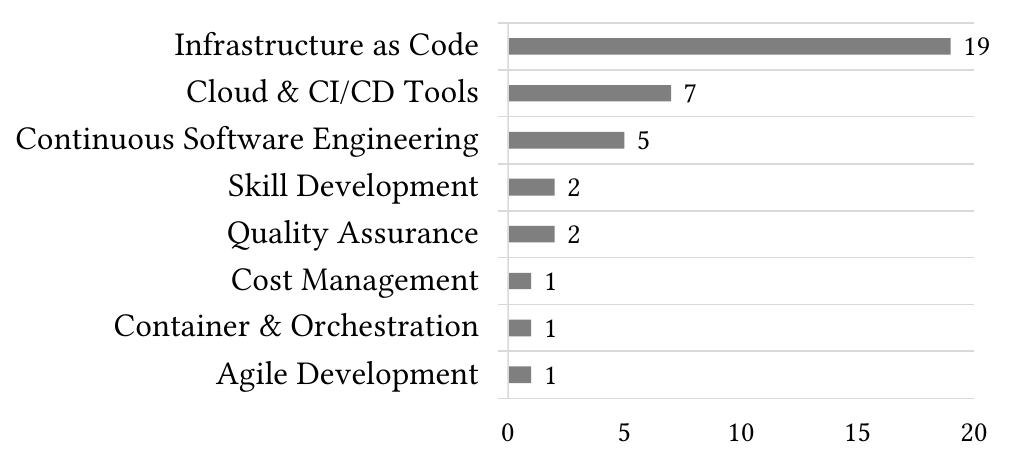} 
\caption{Responded topics for Cloud Infra trend}
\label{fig:survey-q15}
\end{figure}

\nd\textbf{Future Trend of Cloud Infra. (Q15):} The participants opined that cloud infrastructure automation will be a critical part of DevOps and programmers will have to improve their infrastructural skills to keep up. Figure \ref{fig:survey-q15} shows responded topics. 

\begin{tcolorbox}[flushleft upper,boxrule=1pt,arc=0pt,left=0pt,right=0pt,top=0pt,bottom=0pt,colback=white,after=\ignorespacesafterend\par\noindent]
\noindent\textbf{RQ$_4$ What additional insights about DevOps challenges can we get from open-ended responses of survey participants?} 45\% of participants consider continuous integration as the most critical component of DevOps practice. Though, none of 23 topics found in RQ$_1$ refers to monitoring tools, 25\% participants considered continuous monitoring to be the most important DevOps process. The respondents emphasized lack of skilled professionals as a major problem and only 5\% organizations offer formal training. 
\end{tcolorbox}

\section{Implications of Findings}\label{sec:implicatons}
Our study findings can guide the following DevOps stakeholders: 
\begin{inparaenum}
\item {DevOps Vendors} to develop more usable DevOps tools,
\item {Organizations} to allocate more resources to train the DevOps practitioners, 
\item {Solution Architect} to decide on tools like Kubernetes for container and orchestration, 
\item {Delivery/Operations Manager} to use the popular automation tools and to collaborate with solution architects to devise fallback and recovery plan, 
\item {DevOps Practitioners} to determine learning paths based on DevOps topics, and 
\item {DevOps Researchers and Educators} to develop innovative DevOps technology to increase the rapid adoption and usage of DevOps tools. 
\end{inparaenum} We discuss the implications below.

\begin{figure*}
\includegraphics[scale=.35]{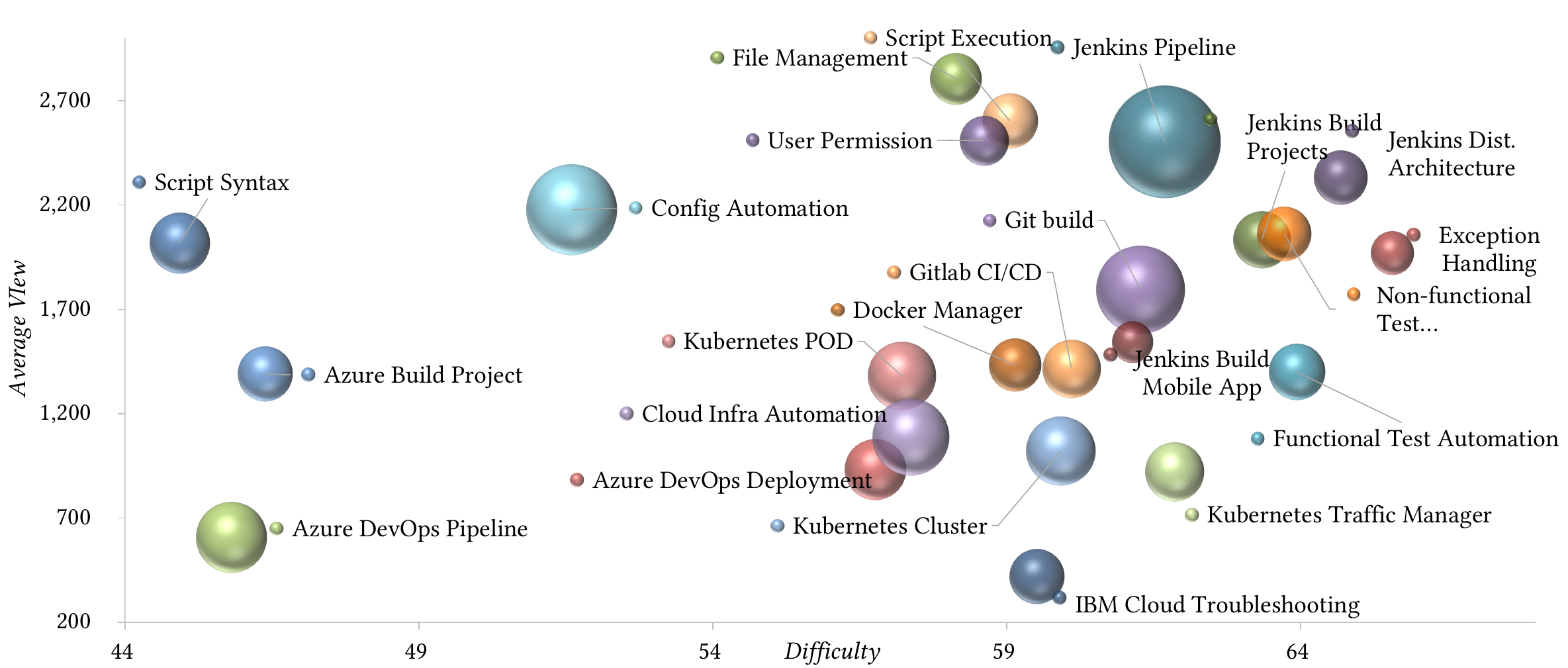}
\caption{DevOps Topics popularity vs. difficulty}
\label{fig:Popularity Difficulty Tradeoff}
\end{figure*}
\nd\bf{\ul{DevOps Vendors.}} In \fig\ref{fig:Popularity Difficulty Tradeoff}, we show a trade-off between the popularity and difficulty of the 23 DevOps tools and topics in our dataset. The more right a topic is the more difficult it is. The higher a topic is at the top, the more popular it is. The size of each bubble represents the total number of questions. Jenkins Pipeline is one of the most difficult topics and also among the most popular ones. Therefore, Jenkins vendors need to make the pipeline creation process and techniques in Jenkins easier to use. Similarly Cloud CI/CD developers can also get insight as indicated by experts in the qualitative survey \textit{"To make the data centres and systems entirely configurable is going to be the demand for the next generation devOps engineers where they will make all the cloud infrastructure configuration settings possibly only by the code they will write"}.  

\nd\bf{\ul{Software Organizations.}} While inquiring about the difficulty of DevOps topics to our survey participants, we find that very few (5\%) of them have formal support for training from their companies. This is irrespective of our survey participants coming from small, medium and large organizations. Therefore, software organizations can allocate more training for DevOps practitioners. Formal and frequent training are important, given many practitioners may not always find solutions to their problems in forums like SO (having 58\% unsolved questions). 

For conducting DevOps related training, one important part is preparing the syllabus of training. Our paper provides a complete mapping of all DevOps related concepts and hence, can help a lot as a training outline. Moreover, in all aspect of DevOps, we also listed down most notable tools under section \ref{sec:study-results}. As we found from the practitioners' survey that tools are very important part of DevOps, organizations can also run such tool specific training.

\nd\bf{\ul{Solution Architects.}} From our study, active communities in DevOps topics can be found and architects can consider community support factor for selecting technologies. For example, three topics of Kubernetes (pod, cluster, traffic management) have average difficulty level of around 60\% which is just above average difficulty level (59\%) of all topics (see \fig\ref{fig:Popularity Difficulty Tradeoff}). We find that these topics have one of the fastest accepted answers from the community which is around 106 to 127 hours on average (among top 5 topics with fastest resolution). Because of this active community support, solution architects can more confidently use Kubernetes in their solution design. On the other hand, comparing between Jenkins and GitLab for continuous integration, a team may reconsider choosing GitLab which has slowest community support (250 hours of waiting period for acceptable answers)

Moreover, our study can be used as an index for tools and techniques by solution designers. Solution architects who are designing a new system can identify important DevOps practices from Figure \ref{fig:DevOps_Topics} for building an end-to-end automated pipeline and can find detail tools in section \ref{sec:study-results}. For example, they may be already aware of configuration automation (using Ansible, Puppet etc.), but they can also instantly learn about cloud infrastructure automation and related tools (terraform, cloud formation etc.) which are an emerging technology.

\nd\bf{\ul{Delivery/Operation Managers.}} The study can be used for planning DevOps adoption process in organizations. For example, in the \fig\ref{fig:Popularity Difficulty Tradeoff}, Configuration Automation is represented by the second biggest bubble as its post count (16,000) is more than 2.5 times of the median post count (6,000). The posts are also frequently resolved, as the bubble lies at the left of the chart. Using insight of this quantitative result, managers who want to adopt or initiate DevOps practices can start by automating configurations with related popular tools such as Ansible, Chef, Puppet, etc.

Our study can also help managers in making realistic delivery plans. Apparently, it may seem that using DevOps tools will automate the delivery cycle and ensure faster production rollouts. However, our study shows that setting up DevOps tools are not always easy and developers often struggle with a lot of difficulties throughout the pipeline.

For example, if a team is suddenly stuck with mobile application build through Jenkins, our paper should be able to guide a manager that this is a very difficult situation to handle, since almost 61.1\% SO posts does not have any resolutions and even if they get lucky, the median wait time can be as long as 10 days (248 hours to be precise), as listed in Table \ref{tab:topicPopularityDifficulty}. Hence, managers studying our paper should be able to anticipate the roadblocks and can help their team with proper planning and necessary resources.

\nd\bf{\ul{Developers.}} Our qualitative analysis has identified the need for Skill Development in DevOps. A respondent specifically emphasized on the infrastructure skill of developers by saying \textit{"As it [cloud infra automation] is mostly used by people having programming exposure, the lack of system and networking knowledge in this set of people is a major problem in my opinion."}. This also validates the quantitative finding on Infrastructure as Code category. Topics such as file \& permission management, script execution have less number of posts, but their average view count is highest among all topics (top side in the \fig\ref{fig:Popularity Difficulty Tradeoff}). Therefore, aspiring developers can start learning tools related to these topics and then move to more advanced topics like Jenkins pipeline.

In this paper, we have reported that most knowledge transfer regarding DevOps happens ad-hoc without any formal training. It is only to the best interest of the developers that they learn about the concepts and tools online by themselves, since we have identified that only 10\% of developers take online DevOps courses to improve their skill.

\nd\bf{\ul{Educators and Researchers}} can create tutorials and documentation for the most difficult topics as shown in \fig\ref{fig:Popularity Difficulty Tradeoff}. Researchers can also investigate the reason of delay in getting acceptable solutions in SO posts. In addition, DevOps research could take cues from crowd-source and big-data research to develop techniques that can
automatically find acceptable answers to unanswered questions, e.g., by recommending acceptable answers to a
question~\cite{Asaduzzaman-FindingAnswersToUnansweredQuestions-MSR2013,BowenXu-AnswerBot-ASE2017}.

\section{Threats to Validity}\label{sec:threats}

\bf{External Validity} threats concern the
generalizability of our findings. We focused on SO, which is one of the largest
and most popular developers' Q\&A Web sites. Yet, our findings may not
generalize to other Q\&A Web sites. We only considered questions and accepted answers in our topic modeling. Our approach is consistent with previous work that used topic modeling on SO data~\cite{Zhang-AreCodeExamplesInForumReliable-ICSE2018,Yang-QueryToUsableCode-MSR2016,Terragni-CSNIPPEX-ISSTA2016,Ren-DiscoverControversialDiscussions-ASE2019,Asaduzzaman-FindingAnswersToUnansweredQuestions-MSR2013,Ponzanelli-ImprovingSOPosts-ICSME2014,Bagherzadeh2019,Rosen-MobileDiscussionSO-EMSE2016}. 

\bf{Internal Validity} threats concern experimental
bias and errors while conducting the analysis. In particular, in our study, we
manually labeled the topics. To reduce any bias in this labeling, two different
authors separately labeled the topics and then another author, who is a domain
expert, validated the labeling. The three authors discussed any conflict and
resolved them via discussions. Thus, we believe that we reduced labeling bias to
an acceptable minimum.  

\bf{Construct Validity} threats relate to potential
errors that may occur when extracting data about DevOps-related discussions. We
collected all SO posts labeled with one or more tags related to DevOps, i.e., 19
different tags. We created the list of tags using state-of-the-art
approaches~\cite{YangLo-SecurityDiscussionsSO-JCST2016,Bagherzadeh2019} and by manually verifying the tags. Threats to construct validity also pertain to the difference between theory, observation, and
results. Our use of metrics to measure popularity and difficulty fall under such
threats. Yet, we used metrics that were used in previous
works~\cite{abdellatif2020challenges,Bagherzadeh-BigdataTopic-FSE2019}, thus mitigating the risk of wrong measurements. In the questionnaire survey, asking people to indicate a context free difficulty level for a broad category of topics may introduce some inaccuracy to measure these perceptions, since difficulty level of a task varies from technology to technology in some cases. However, the participants are likely to consider majority applications while responding to such queries and some peripheral cases may be overlooked in absence of better alternative method. 
\section{Related Work} \label{sec:related-work}
\nd\bf{\ul{DevOps Definition.}} Aiello et al. defined DevOps in \cite{Aiello-AgileDevOps-2016} as \textit{DevOps is a set of principles and practices intended to help development and operations collaborate and communicate more effectively.} In a literature survey on DevOps \cite{Leite-DevopsSurvey-ACM2020}, Leite et al. identified four major categories of DevOps concepts, namely, Process, People, Delivery, and Runtime using a systematic analysis on 50 'core' DevOps publications till 2019. While process and people are relevant from the management perspective, delivery (development related) and runtime (operation related) are associated from the engineering perspective.

\nd\bf{\ul{Case and Industry Studies on DevOps.}} Besides the academic researches, there has been a industry wide survey and state of DevOps report published by Puppet since 2011 \cite{website:devops-state-2021}. They surveyed 2,400 industry-people in 2020  \cite{website:devops-state-puppet} and in 2021, the report investigates team interactions and recommends actions toward "breaking down the middle". There are few more studies on DevOps definition, literature survey, and case studies on DevOps practices and adoption \cite{Jabbari-devops-2016, luz2019adopting, lwakatare2015dimensions, Lwakatare-DevOpsInPractice-IST2019, smeds2015devops, Senapathi-devops-case-2018, Guerriero-iac-8919181} and as well as a literature survey on DevOps along with qualitative interview with six organizations on their practices \cite{erich2017qualitative}

\nd\bf{\ul{SE Research on DevOps.}} There are few research works that focused on different aspects of DevOps practice. The studies on Continuous Software Engineering ~\cite{Mansooreh-EASE2020}, Release Engineering \cite{openja2020analysis}, Dockerfiles ~\cite{Docker-MSR2017} and Docker ~\cite{Docker-ESEM2020} provide insight on the specific processes or tools of DevOps engineering. The research ~\cite{Docker-ESEM2020}  presents a study of LDA-based topic modeling on the community discussion related only with Docker on topics. The work of ~\cite{Docker-MSR2017} concentrates on the quality and evolution of Dockerfiles.

\nd\bf{\ul{Research on Continuous Software Engineering (CSE).}} A closely relevant work by ~\cite{Mansooreh-EASE2020} performs empirical study on CSE. As outlined by the authors of \cite{bosch2014continuous, fitzgerald2014continuous, fitzgerald2017continuous} and also in the background section, CSE has few common processes (continuous planning, continuous testing, CI/CD etc.) as DevOps, but DevOps is a much broader concept covering organizational change management and engineering practices \cite{Leite-DevopsSurvey-ACM2020}. 

In the empirical study on CSE, the authors also explicitly attempted to avoid Stack Overflow posts concerning the tools of DevOps, still 12 out of 32 topics, identified by LDA-based topic modeling, were directly related with specific DevOps tool. At the same time, 186 out of 200 posts qualitatively analyzed by the study were directly related with DevOps tools. Despite the researchers' attempt to avoid DevOps tool, their inevitable presence in topics and posts refers that tools are a significant part of DevOps. Though our study apparently overlaps with that research, our empirical study comprises of 13 times larger dataset and 23 consolidated topics focused all around DevOps. Moreover, we conducted an external practitioner survey to validate the empirical study and collected qualitative insights on DevOps on people and process concepts. 

The major technical topics missing in the CSE study is automation. The paper did not report any topic regarding automated testing, configuration automation or cloud infrastructure automation which comprises 18\% of DevOps post we found. Secondly, their paper missed leading configuration management tools such as Ansible, Puppet and major cloud infrastructure automation tool Cloud Formation. The study did not mention Kubernetes and did not discuss any topic regarding orchestration which consists of 13.4\% of DevOps topics we found. Furthermore, their analysis missed Infrastructure as Code, a very important concept of DevOps having 28.9\% of discussions in Stack Overflow.

\nd\bf{\ul{Study on Software Release Engineering.}} Another recent paper \cite{openja2020analysis} performs a similar empirical study on release engineering questions in Stack Overflow, using LDA-based topic modeling techniques. As explained in the background section, Release engineering can be seen an important but smaller subset of DevOps. This is also evident since the topics found in the release engineering paper \cite{openja2020analysis} are clearly separated from most of the DevOps topics found in our empirical study. 

The release engineering paper reports 'Mobile Deployment' as a topic, whereas our paper discussed 'Build Mobile App' as a DevOps topic. It seems that their focus was more on the delivery side concerns such as deployment or build failure, whereas we found unique DevOps topics covering full build lifecycle from Git Build, Build Project or end to end Jenkins Pipeline. Moreover, Jenkins was not considered as any standalone topic there and continuous integration comprised of only 3.3\% discussion, whereas we reported a massive 21.4\% SO posts only regarding Jenkins which is the most prominent continuous integration tool.

In spite of focusing on release engineering, their paper did not report any topic regarding Kubernetes, the state-of-the-art orchestration tool, whereas we reported that Kubernetes holds a major part, around 13.4\%, of all DevOps questions. In release engineering, the authors mostly presented traditional testing and infrastructure or configuration management without automation, whereas in DevOps concept, we found that 18\% of posts are more concerned in automation aspects such as configuration automation (8.6\%), cloud infra automation (6.2\%), functional test automation (3.2\%).

Since management perspective is one core concept of DevOps, our paper discussed organizational issue such as agile development process and developers' training for DevOps skill development, whereas such perspective is missing in the release engineering study.

All the above studies are limited to the building, integration, and deployment process and to the best of our knowledge there is yet to be a study on the overall process, people, delivery and runtime aspects of DevOps. The summary of the differences along with our research contribution is provided in the Table \ref{tab:contribution1} and Table \ref{tab:contribution2}. In this paper, we report a mixed methods study on holistic view of DevOps concepts and on challenges faced by engineers by analyzing 174K SO posts and by conducting a survey of 21 DevOps professionals in the industry.

\begin{table}[]
\hspace{-5mm}
    \centering
    \caption{Comparison with different DevOps related study types}
     \resizebox{\columnwidth}{!}
  {%
    \begin{tabular} 
    {p{0.15\linewidth} |  p{0.3\linewidth} |  p{0.10\linewidth} | p{0.70\linewidth}}
    \toprule
       \textbf{Study Type} & \textbf{Previous Study} & \textbf{DevOps Topic} & \textbf{Research Contribution} \\ \midrule
        Definition and Literature Survey & "DevOps dimension by Lwakatare et al. \cite{lwakatare2015dimensions}
DevOps definition by Aeillo et al. \cite{Aiello-AgileDevOps-2016}
Literature survey by Leite et al. \cite{Leite-DevopsSurvey-ACM2020}
Mapping study by Jabbari et al. \cite{Jabbari-devops-2016}" & DevOps Definition & Our research is based on the definition provided by these research outcomes. Specifically, we performed open coding of our survey respones on the basis of DevOps aspects (management, engineering, people and process) defined by Leite et al. \\ \hline
        Case studies & Case study by Luz et al. \cite{luz2019adopting}, by Smeds et al. \cite{smeds2015devops}, by Senapathi et al. \cite{Senapathi-devops-case-2018}, by Lwakatare et al. \cite{Lwakatare-DevOpsInPractice-IST2019}, by Erich et al. \cite{erich2017qualitative}, by Guerriero et al. \cite{Guerriero-iac-8919181}. & Adoption and Practice & These are specific studies performed on the case specific struggles faced by the companies. However, in our research, we also cover the challenges in the industry through a qualitative survey. Our survey has provided insight on industry wide challenges on skill shortage, and lack of organizational training. \\ \hline
    \multirow{8}{*}{SE Research} & Research by Schermann et al. \cite{Docker-MSR2017}, by Haque et al. \cite{Docker-ESEM2020} & Docker Discussion & Specific studies performed on Docker tool only. Our research finds that Docker discussion comprises of only 2.7\% of all DevOps discussions. \\ \cmidrule{2-4}
       {} & Study by Rahman et al. \cite{rahman2018questions}  focuses mainly on Puppet, a major tool of Infrastructure as Code (IaC) & IaC Discussion & In our research, we found that there are other IaC tools such as TerraForm, CloudFormation which are gaining popularity in discussion (having 6.2\% of questions in SO). Overall, Puppet and Chef combined is discussed 4.6\% times among all DevOps topics in Stack Overflow. \\ 
       \bottomrule
    \end{tabular}%
    }
    \label{tab:contribution1}
\end{table}

\begin{table}[]
\hspace{-5mm}
    \centering
    \caption{Additional Contribution to CSE and RE domain related researches}
     \resizebox{\columnwidth}{!}
  {%
    \begin{tabular} 
    {p{0.15\linewidth} |  p{0.3\linewidth} |  p{0.10\linewidth} | p{0.7\linewidth}}
    \toprule
       \textbf{Domain} & \textbf{Previous Study} & \textbf{DevOps Topic} & \textbf{Research Contribution} \\ \midrule
       Continuous Software Engineering & Research by Zahedi et al. \cite{Mansooreh-EASE2020} focused on the engineering process aspects of DevOps only. They explicitly tried to avoid the DevOps related tools from the dataset by excluding related tags in Stack Overflow. & CI/CD Discussion & Our empirical dataset is 13 times larger than this study. Moreover, this study has limitation for excluding the tools related tags in SO. Also, unlike ours, this study does not provide grouped categories of the 32 CI/CD related topics. As a result, this is very specific and granular level research focusing only on the CI/CD process. However, in our study, we have grouped 23 DevOps topics under 4 major categories.  Also, we found that CI/CD tools constitutes of 33.2\% of all DevOps discussions. So, topic coverage wise this study focuses on less than one third of our research area and the topics are very granular. \newline
       The study did not cover infrastructure as code (cloud and configuration automation), orchestration (Kubernetes), and automated testing concepts which consists of almost 48.6\% DevOps related discussion.
       
       \\ \hline
       Release Engineering & Relevant study by Openja et al. \cite{openja2020analysis} focused on the release engineering steps related discussions in Stack Overflow till December 2017. & Release Engineering Discussion & \textbf{Firstly}, the focus area of this study only covers a portion of total DevOps concept. For example, because they only mined ReLeng topics, Continuous Integration came in 3.3\% of their posts whereas we found that Continuous Integration tool Jenkins was discussed in 21.4\% of all DevOps posts. They found highest posts in Software Testing (5.2\%) since they only considered the Release phase of the SDLC, but in total DevOps we found that Quality Assurance (functional and non-functional) posts are more frequent (6.3\%). \textbf{Secondly}, since their study is based on almost 3 years old data (December 2017 SO dump vs ours July 2020 SO dump), many of the latest trends have been missed; for example, we found that Infrastructure as Code is discussed in 28.9\% of posts, whereas according to their findings it was only 11.1\% of ReLeng discussion. \textbf{Finally}, we extended our research on the people, organizational, and process aspects of DevOps \cite{Leite-DevopsSurvey-ACM2020} by conducting a qualitative explanatory survey and identified skill, and training related hurdles in DevOps whereas their study only performed empirical data analysis. \\ \bottomrule
          
    \end{tabular}%
    }
    \label{tab:contribution2}
\end{table}

\section{Conclusion} \label{sec:conclusion}
In this mixed methods empirical research, we have applied topic modeling on 174K Stack Overflow posts that contain DevOps related discussions. We have found 23 DevOps topics grouped into four categories. The topic category `Cloud \& CI/CD Tools' contains the most number of topics (10) which cover 48.6\% of all questions in our dataset, followed by the category Infrastructure as Code (28.9\%). The file management and script execution topics from the category Infrastructure as code are the most popular (with most views), while infrastructural Exception Handling, and non-functional test automation have the highest percentage of unsolved questions. We validate the quantitative study by and provide qualitative analysis on DevOps challenges by conducting a survey among 21 industry professionals. Our findings can be used by the vendors of DevOps tools to address the issues reported by practitioners; by architects and managers for system design and DevOps adoption decisions. In future, large scale exploratory survey can be conducted to gain insights for the solution on DevOps challenges.

\bibliography{references}

\end{document}